\documentclass[12pt,preprint]{aastex}

\def\today{\rightline{\ifcase\month\or
        January\or February\or March\or April\or May\or June\or
        July\or August\or September\or October\or November\or December\fi
        \space\number\day, \number\year}}

\begin{document}
\title{Probing Inward Motions in Starless Cores Using
The HCN $J = 1 - 0$ Hyperfine Transitions : A Pointing Survey
Toward Central Regions}
\author{Jungjoo Sohn\altaffilmark{1,2}, Chang Won Lee\altaffilmark{2},}
\author{Yong-Sun Park\altaffilmark{1}, Hyung Mok Lee\altaffilmark{1},}
\author{Philip C. Myers\altaffilmark{3}, and Youngung Lee\altaffilmark{4}}

\altaffiltext{1}{Astronomy Department, School of Physics and
Astronomy, Seoul National University, Shillim-dong, Kwanak-gu,
Seoul 151-749, Korea}
\altaffiltext{2}{International Center for Astrophysics, Korea
Astronomy \& Space Science Institute, 61-1 Hwaam-dong, Yusung-gu,
Daejeon 305-348, Korea}
\altaffiltext{3}{Harvard-Smithsonian
Center for Astrophysics, 60 Garden Street, MS 42, Cambridge, MA
02138, USA}
\altaffiltext{4}{Korea Astronomy \& Space Science
Institute, 61-1 Hwaam-dong, Yusung-gu, Daejeon 305-348, Korea}
\clearpage
\begin{abstract}
We have carried out a survey toward the central regions of 85
starless cores in HCN $J=1-0$ to study inward motions in the
cores. Sixty-four cores were detected with HCN lines. The infall
asymmetry in the HCN spectra is found to be more
 prevalent, and more prominent than in any other previously used infall
tracers such as CS $J=2-1$, $\rm DCO^+$ $J=2-1$, and $\rm N_2H^+$
$J=1-0$. We found close relation between the intensities of the
HCN and $\rm N_2H^+$ lines. This implies that the HCN is not much
depleted in the central regions of the cores. In some cores, the
HCN spectra show different sign of asymmetry from other molecular
lines. A few cores show various signs of asymmetry in individual
HCN hyperfine lines. The distribution of the velocity shift
$\delta V$ of the HCN profiles with respect to the systemic
velocity of the optically thin tracer is found to be more shifted
toward bluer side than those of other infall tracers, indicating
that the HCN traces inward motions more frequently. The $\delta V$
distribution of each HCN hyperfine line for all sources is
similar. Moreover the $\delta V$ values obtained from different
HCN hyperfine lines for each source are nearly similar. These may
mean that most of starless cores are in similar kinematic states
across the layers of the cores. We identify 17 infall candidates
using all available indicators such as the velocity shift $\delta
V$ and the blue to red peak intensity ratio of double peaked
profiles for HCN $J=1-0$, CS $J=2-1$, $J=3-2$, DCO$^+$ $J=2-1$,
and $\rm N_2H^+$ $J=1-0$. Four of them, L63, L492, L694-2, and
L1197 are found to show higher blue to red ratio in the HCN
hyperfine line along the lower opacity, suggesting that infall
speed becomes higher toward the center.
\end{abstract}

\section{Introduction}

Starless cores are dense molecular condensations without Yong
Stellar Objects (YSOs) (Beichman et al.  1986; Lee \& Myers 1999,
hereafter LM99; Di Francesco et al. 2006). Some of the starless
cores are called ``prestellar cores'' which show more evolved
features of the starless cores  such as  relatively higher $\rm
H_2$ column density, more compact density distribution, 
deuterium enrichment, more pronounced CO depletion, and inward
motions (Crapsi et al. 2005). Because of this, they are believed
to be in a stage close to form a star. On the other hand there are
other types of starless cores which do not show any good hints of
star formation activity. It seems that not every starless core is
destined to deliver a baby star. In most cases, however, the
starless cores are believed to be one of the best targets to study
the initial conditions of star formation and its related physical
and chemical processes (Benson \& Myers 1989; Ward-Thompson et al.
1994; Shirley et al. 2000; Andr\'{e}, Ward-Thompson, \& Motte
1996).

Inward motions are one of the key processes to study physical and
kinematical status of the starless core, and to test models of
star formation. The first evidence of the existence of such
motions in starless cores was discovered from a study of the
spatial distribution of the ``spectral infall asymmetry'' toward a
starless core L1544, the spectral line combination of a self
absorbed profile with double peaks of a blue component being
brighter than red one in an optically thick tracer and a single
peak profile in an optically thin tracer (Tafalla et al. 1998).
The infall asymmetry found in L1544 was spatially extended ($\sim
0.1$ pc) with its deduced infall speed of up to 0.1 km s$^{-1}$.
Lee, Myers \& Tafalla (1999, LMT99 hereafter) found from a
pointing survey of starless cores in CS $J = 2 - 1$ and $\rm
N_2H^+$ $J = 1 - 0$ lines that the inward motions are the most
dominant kinematics in starless cores with a density higher than
about $10^{4}$ $\rm cm^{-3}$. From their follow-up mapping
observations of starless cores in the same lines, Lee, Myers \&
Tafalla (2001, LMT01 hereafter) have showed that the infall
asymmetry found in L1544, spatially extended up $0.2$ pc with the
line of sight speed of order 0.1 km s$^{-1}$, is a general feature
in the collapsing starless cores. The extended nature of the
infall asymmetry has an implication that the inward motions might
exist before the formation of a central star. This is inconsistent
with the inside-out model by Shu (1977) (Tafalla et al 1998). Lee,
Myers, \& Plume (2004, LMP04 hereafter) reported that several
cores may have higher infall speed in the inner denser regions
from the comparison between the CS $J = 3 - 2$ and CS $J = 2 - 1$
line surveys. Recently, Williams, Lee, \& Myers (2006) showed
spectral signatures of rotation and collapse in two starless
cores, L1544 and L694-2, by comparison of interferometric map
observations in $\rm N_2H^+$ $J = 1 - 0$. They found that the less
evolved core, L694-2, has smaller rate of velocity increase toward
the center than the more evolved core, L1544, indicating that
development of the spatial structure of the infall speed may be
related to the evolutionary stage of the starless cores. These
observational results on the inward motions in starless cores will
be useful to improve the collapsing models explaining the features
of inward motions in the early stage of star formation (e.g.,
Whitworth \& Ward-Thompson 2001, Ciolek \& Basu 2000, and Myers
2005).

One obstacle in this kind of study is a possible depletion of the
infall tracer in the dense central regions of the cores (Caselli
et al. 1999; Tafalla et al. 2002; Alves, Lada \& Lada 2001;
Charnley 1997). Leger (1983) and Bergin \& Langer (1997) have
shown that in a cold core denser than a few $10^{4}$ $\rm
cm^{-3}$, polar molecules such as CS are easily depleted by being
adsorbed on the dust grains. This property may often limit the
study of inward motions occurring in the nuclear regions of the
cores. For example, the typical size of CS depleted region in
starless cores observed by Tafalla et al. (2002) is about 0.1 pc
which is almost comparable to the size of the region showing
extended infall asymmetry. This implies that most of the blue
skewed CS profiles seen in the infalling cores may not be formed
from the central region, but mainly from envelopes of the cores.
Therefore, CS line may not be the best to trace infall motions
occurring at the nucleus of the cores.

This study introduces a survey of infall motions toward starless
cores with HCN $J = 1 - 0$ line. Though the intensity of HCN is
weaker than that of other transitions, this line has an advantage
of tracing the central motion more sensitively than CS (e.g., Park
et al. 1999). Our follow-up mapping observations of several
starless cores in HCN $J = 1 - 0$ (Sohn et al. in preparation)
indicate a similar distribution to that of $\rm N_2H^+$ which
usually traces the density distribution of inner part of the cores
(Tafalla et al. 2006, Di Francesco et al. 2006). Aikawa et al.
(2005) have shown that HCN $J = 1 - 0$ does not suffer from
significant depletion effect in the central density of the core
less than 10$^{5}$ cm$^{-3}$. The calculation of the evolution of
the HCN abundance by Lee et al. (2004) indicates that HCN is
slightly depleted in a very small (of order $10^{-3}$ pc) region
near the center of the core, at the moment right near the stage of
the star formation, but not as much as the CS molecule. This
``small'' depletion is seen in two starless cores, L1498 and
L1517B (Tafalla et al. 2006).

Another advantage of HCN is the presence of three hyperfine lines
with different relative opacities of F(0-1) : F(1-1) : F(2-1) = 1
: 3 : 5 under the LTE condition. Because the hyperfine lines have
almost equal excitation temperature, these are expected to trace
somewhat different regions mainly depending on their given optical
depths. HCN line can be useful to study spatial variation of
inward motions along the line of sight of the cores.

In this paper we report our systematic survey of 85 starless cores
with HCN $J = 1 - 0$ hyperfine lines to investigate the occurrence
and pattern of inward motions. In \S 2, we explain observations of
the starless cores. In \S 3 we describe the detection statistics,
spectral features and analysis of the observed HCN lines.
Implications of the observational results are discussed in \S 4.
Finally, we summarize our results in \S 5.

\section{Observations}

We have conducted single-pointing observations for 85 starless
cores in HCN $J = 1 - 0$ with the 13.7 m telescope of the Taeduk
Radio Astronomical Observatory (TRAO) in Korea. Observations were
carried out during January, February, November and December in
2002 and January to May in 2003. We used the SIS receiver and
autocorrelation spectrometer with 10 KHz (corresponding to a
velocity resolution of 0.033 km s$^{-1}$ at 88.6 GHz) or 20 KHz
resolution. We adopted a frequency switching mode with frequency
throw of 5 or 8 MHz. Calibration was frequently done by measuring
 the ambient temperature. System temperature during the observations
was typically between 400 - 600 K for single sideband. Pointing
accuracy during the observation was within $\pm$ 10 \arcsec (rms).
HPBW of TRAO telescope is 64\arcsec, and beam efficiency is 50$\%$
at 86 GHz frequency (Roh \& Jung 1999). We also observed 10
starless cores in $\rm H^{13}CN$ $J = 1 - 0$ line with similar
instrument configuration and observing mode between April and May,
2003. The system temperature during the observation was between
280 and 330 K.

Targets were selected primarily based on the strengths of
N$_{2}$H$^{+}$ $J = 1 - 0$, CS $J = 2 - 1$ (LMT99, LMT01), and
HCO$^{+}$ $J = 1 - 0$ (Lee et al. 2007 in prep.) lines. Table 1
lists the names and positions of the sources and the observed line
parameters. All sources are known to be nearby (within a few
hundreds pc), dense (around 10$^{4-5}$ cm$^{-3}$), compact (r
$\approx$ 0.05 $\sim$ 0.35 pc) (LMT99), and have narrow line
widths ($\Delta$V$_{\rm FWHM}$ of N$_{2}$H$^{+}$ $\sim 0.2$ - $0.4
$ km s$^{-1}$, LMT99).

For a quantitative analysis of inward motions using HCN lines, it
is necessary to determine the accurate frequency of HCN. A usual
way to determine the frequency of the HCN is to observe a source
showing a Gaussian shape in both HCN and other reference lines
such as C$^{18}$O $J = 1 - 0$ or N$_{2}$H$^{+}$ $J = 1 - 0$ whose
frequencies are already accurately known. However, most of HCN
profiles turned out to be highly affected by high optical depths,
showing either double-peaked or skewed features, making it
difficult to decide its frequency from the observations.  So, we
use the value of the Cologne Database for Molecular Spectroscopy
giving the best estimation of HCN frequency with the lowest
measurement uncertainty of $\rm \sim 1~KHz$ (M\"uller et al. 2005)
which is smaller than the spectrometer resolution by a factor 10
$\sim$ 20. The value for HCN $J = 1 - 0$ hyperfine lines are as
follows; 88630.4157 ($\pm 0.001$) MHz for ($F$=1$-$1), 88631.8473
($\pm 0.001$) MHz for HCN ($F$=2$-$1), and 88633.9360 ($\pm
0.001$) MHz for ($F$=0$-$1). The reduction of the spectral line
data was performed using the SPA developed by FCRAO and the CLASS
reduction software (Buisson et al. 1994).

\section{Result}
In total sixty four sources out of 85 targets are detected in HCN
$J = 1 - 0$ and eight of ten observed sources are detected in $\rm
H^{13}CN$ $J = 1 - 0$.
The observed line parameters of the sources are listed in Table 1.
The antenna temperature (T$_{A}^*$) of the HCN line ranges
0.17 $\sim$ 1.50 K for ($F$=2$-$1). The $\rm H^{13}CN$ is usually
very weak : $\rm T_{A}^{*} = $ $0.06\sim 0.23$ K. Note that most
of the detected sources are observed in the HCN line with high S/N
($>$ 10). The profiles of the observed lines for all detected
sources are displayed in Figure 1.

\subsection{The HCN Spectra}

One of the most noticeable features seen in Figure 1 is that the
HCN profiles hardly show any clear single Gaussian feature.
Instead, most of the HCN $J = 1 - 0$ spectra show asymmetric
feature such as double peaks with a dip or a single peak with
shoulder. The peak velocity of the optically thin tracer $\rm
N_2H^+$ $J = 1 - 0$ line is located close to a dip between the asymmetric profiles. 
The estimated velocity distance between the peak velocity of $\rm
N_2H^+$ and the dip of HCN profile for 17 sources showing clear dip
in the HCN profiles is $0.1\pm 0.08$ $\rm km~s^{-1}$.
Therefore such a dip is
thought to be caused by self-absorption of emissions from the warm
nucleus by the cool envelope of the cores. It is interesting to
note that the self-absorption dip tends to be deeper at the
hyperfine component with large opacity (HCN, $F$=2$-$1), and at
the same time the infall asymmetry tends to be more clearly seen
at this component. Based on the details line asymmetry, the HCN
spectra can be classified into four groups (See Fig. 2) ;

Group 1: All three hyperfines have two peaks, with a blue peak
brighter than a red peak. It occupies around 9 \% of the detected
sources ( Fig. 2a: L1495AN, L1521F, L1544, L1689B, L492, and
L694-2).

Group 2: One or two of the hyperfines has two peaks, with a blue
peak being brighter than a red peak or a red shoulder component.
About 34 \% of the detected sources fall into this group (Fig. 2b:
L1333, L1355, L1498, L1495A-S, B217-2, B18-1, TMC2, TMC1, L1507A,
CB23, L1552, L1582A, L1622A2, L1622A1, CB45, L1696A, L1696B, L63,
L673-7, L1155C1, L1251A2, and L1197).

Group 3: At least one of hyperfines has two peaks, with a red peak
 being brighter than a blue peak or a blue shoulder. It occupies around
22 \% (Fig. 2c: B5-1, L1399-2, L1521E, L1400A, B18-5, CB22,
L1527B1, L1517B, L1512, L1621-1, L134A, L1704-1, L204F, and
L1049-1).

Group 4: Red and blue asymmetric profiles are mixed in three
hyperfines. This group has around 33 \% in total share (Fig. 2d:
L1521B2, L1521B, B18-3, L1517C1, L183, L1709B2, L204C-2, L234E-S,
L462-1, L462-2, L429-1, L490, CB130-1, L648-1, L1041-2, L922-1,
L1049-2, L1155C2, L944-2, L1251 and CB246-2).

L1633 does not fall into any of the above mentioned groups because
its profiles have broad peaks without clear double peaks at two
hyperfines. The sources in groups 1 and 2 are referred to having
blue asymmetry as the optically thin $\rm N_2H^+$ line show a
single peak located near the dip or shoulder of the HCN profiles.
These groups occupy around 43 \% of the detected sources. On the
other hand the sources in group 3 have red-asymmetry.
 The cause of such a red asymmetry is not clear: it could be due
 to expansion as opposed to the collapse.

We note that there are also the sources that have some interesting
features in HCN spectra. There are two distinct groups. There is a
group of sources having extraordinarily broad line width, almost
twice of (FWHM of $\sim$ 0.8 km s$^{-1}$) those ($\sim$ 0.4 km
s$^{-1}$) of typical starless cores (See Fig. 2e). These are B5-1,
L1582A, L1633, and L204F ($\sim 6$ \% of the detected 64 sources).

Another interesting group is the sources showing anomalous
intensity ratio among the hyperfine lines. If the sources are in
LTE and the optically thin condition, the relative intensities of
the HCN $J = 1 - 0$ hyperfine transitions should have the ratio of
1 : 3 : 5 for $F$=0$-$1 : 1$-$1 : 2$-$1. However, L1333, L1621-1,
L1527B1, L1633, L1696B, L1709B2, L429-1, L462-2, L492-2, L694-2,
L922-1, CB130-1, and L1197, have almost similar intensities
between $F$=0$-$1 and $F$=2$-$1 lines (See Fig. 2f). Furthermore
the intensity of $F$=0$-$1 hyperfine line in L694-2, CB130-1, and
L1197 is even stronger than that of $F$=2$-$1.  
About twenty percent of the detected sources shows anomalous intensity ratio among
the hyperfine lines. This intensity ratio anomaly is not quite
well understood. One of plausible explanations is that the
$F$=2$-$1 hyperfine line which have higher opacity and therefore
trace the regions where the excitation temperature is lower, while
the $F$=1$-$0 lines which have much lower opacity and those trace
deeper regions where the excitation temperature is higher, as
Cernicharo et al. (1984) and Fuller et al. (1991) have pointed
out.

\subsection{Comparison of the HCN Spectra with other spectra}

The systematic surveys for searching for inward motions in
starless cores have been carried out in several molecular lines
such as CS $J = 2-1$ (LMT99), $J = 3-2$ , and DCO$^+$ $J = 2-1$
(LMP04). Compared with these lines for over 50 sources, the
spectra of HCN are found to have following important features.

The first distinctive feature of the HCN line is the
self-absorption. The HCN line seems to be the most opaque among
the tracers so that the double-peaked shape of the spectrum is
more often seen in HCN (for more than 20 sources) than any other
tracers (about 10 sources in CS $J = 2-1$ and $J = 3-2$ lines).
Moreover, the self-absorbed dip of the HCN is more pronounced and
the separation between the peaks is wider than those of other
tracers. Fig. 3$-$(a) shows the line for L1544 where the
self-absorbed dip of the HCN line almost reaches the zero
temperature level and the separation between two peaks is the
widest among four different tracers.

The second noticeable feature of the HCN profile is that the
profiles are more asymmetric than other tracers. For example, a
red component in the HCN profile L1498 is almost totally absorbed
while DCO$^+$ $J = 2-1$ is close to Gaussian, and CS $J = 2-1$ and
$J = 3-2$ show a blue peak with red shoulder ({\bf Fig. 3$-$(b)}).
HCN profile in L492 shows the double peaked infall asymmetry while
other tracers show the skewed asymmetry or Gaussian shape. More
than 10 sources show this feature. More asymmetric feature in HCN
is also seen in the red profile. For example, the HCN in L1512
shows an extreme red profile so that the blue component is nearly
self-absorbed out while other lines show either Gaussian
(DCO$^+$), or red profile with slight asymmetry (CS $J = 2-1$, and
$J = 3-2$) (Fig. 3$-$(c)).

The last feature to note in the HCN is that some of the HCN
profiles show different asymmetry from other ones, i.e., blue
profile in HCN and red profile in other tracers, or vise versa.
Furthermore HCN profile itself often shows different asymmetry
among three HCN hyperfine lines. Fig. 3$-$(d) displays an example
for L183 showing the typical red profile in HCN $F$=2$-$1 (and
also in DCO$^+$), but the blue profile in CS $J = 2-1$ and $J =
3-2$ lines. Note that, as shown in Fig. 4, L183 shows a blue
profile in HCN $F$=0$-$1, but red profiles in $F$=2$-$1 and
$F$=1$-$1. L429-1 is also a similar case: red profiles in HCN
$F$=2$-$1 and CS $J = 2-1$ while blue profile in CS $J = 3-2$ and
Gaussian one in DCO$^+$. The HCN lines show blue profiles in
$F$=0$-$1 and 1$-$1, but a red profile in $F$=2$-$1.

\subsection{Analysis of $\delta$V for HCN $J = 1 - 0$ hyperfine transitions }

A shape of each HCN hyperfine line, especially a line asymmetry
frequently seen in HCN, can be well quantified by estimating the
amount by which HCN $J = 1 - 0$ hyperfine spectrum is blue- or
red-shifted with respect to a line center of the optically thin
tracer defined in the following equation (Mardones et al. 1997):

\begin{equation}
\rm \delta V = [V_{HCN(F=i-j)}-V_{N_{2}H^{+}}]/ \Delta
V_{N_{2}H^{+}},
\end{equation}
where V$_{{\rm HCN}( F = i-j )}$ is the HCN peak velocity of each
hyperfine ($F$=i$-$j) component obtained from a Gaussian fit. 

The HCN hyperfine lines of double components or a single peak with a shoulder
were fitted only for a brighter component with a Gaussian 
after masking the fainter part or the shoulder  
to obtain the velocity of the peak component of each line.

The peak velocities of N$_{2}$H$^{+}$ lines, V$_{\rm N_{2}H^{+}}$, and
the FWHM of N$_{2}$H$^{+}$ $J = 1-0$, $\Delta$V$_{\rm
N_{2}H^{+}}$, are taken from LMT99. Note that V$_{\rm N_{2}H^{+}}$
was slightly shifted with respect to more accurate frequency
determined by LMT01. The velocity shifts for HCN lines for 50
sources are calculated and listed in Table 1. Figure 4 shows three
histograms for the distribution of $\delta$V for each hyperfine
component. For comparison, we also show that of $\delta$V$_{\rm
CS}$, which was similarly defined as e.q. (1) for CS lines from
LMT99. Distributions of $\delta$V$_{{\rm HCN}(F = 0-1)}$,
$\delta$V$_{{\rm HCN}(F = 1-1)}$ and $\delta$V$_{{\rm HCN}(F =
2-1)}$ are clearly skewed to the blue side ($\delta$V$_{\rm HCN}$
$<$ 0). Mean values of the distributions are $-0.51 \pm 0.06$ km
s$^{-1}$, $-0.59 \pm 0.08$ km s$^{-1}$, and $-0.41 \pm 0.10$ km
s$^{-1}$, for $F$=0$-$1, $F$=1$-$1, and $F$=2$-$1 respectively.
Note that L429-1 and L183 have different asymmetry among three
hyperfine lines, causing the $\delta$V$_{{\rm HCN}(F = 2-1)}$ to
fairly differ from other $\delta$V values for two other hyperfine
lines. Excluding these two sources, $\delta$V values of three
hyperfines ($-0.49 \pm 0.06$ km s$^{-1}$, $-0.56 \pm 0.08$ km
s$^{-1}$, and $-0.50 \pm 0.08$ km s$^{-1}$, respectively) become
very close. A statistical \emph{t}-test with unknown standard
deviation shows that the probability of drawing the observed
$\delta$V$_{{\rm HCN} (F = i-j )}$ from the zero mean normal 
distribution is far less than 1 \% for all three hyperfine lines.
Therefore the skewed distribution for all hyperfine lines is
statistically very significant. As noted in LMT99, and LMP04, the
skewed $\delta V_{{\rm HCN}(F = i-j)}$ distributions to the blue
can be produced only if the cores are predominantly in inward
motions. Thus we conclude that inward motions of the gas are
prevalent in the starless cores. Note that all $\delta$V$_{{\rm
HCN}}$ distributions are more skewed than those of CS $J = 2-1$
($<\delta V>= -0.13 \pm 0.04 ~ {\rm km~s^{-1}}$, LMT99) and CS $J
= 3-2$ ($<\delta V>= -0.13 \pm 0.03 ~ {\rm km~s^{-1}}$, LMP04),
indicating that the hyperfine lines of HCN better trace inward
motions in starless cores than CS, as already implied in the last
section from more asymmetric characteristics of HCN profiles than
those of any other tracers.

It is not quite clear whether $\rm \delta V$ distribution of
HCN component with lower optical depth is more skewed to the blue
side than that of the HCN component with higher optical depth as
shown in Figure 4.
Rather the mean values of $\delta V$ distributions among all HCN
hyperfine components appear to be similar within the error of the
distribution. 
Moreover, Figure 5 shows correlations among $\delta$Vs of each hyperfine
line of the HCN. As noted above, L429-1 and L183 show different
asymmetry among three hyperfine lines and also greatly affect the
quality of correlation among $\delta $Vs of three hyperfine lines.
Excluding these two unusual sources, the correlation among the
$\delta$Vs of each hyperfine line becomes much better (the
correlation coefficient of the fit becomes better than 0.8).
This may imply that the gas in starless cores has
inward motions in a systematic rather than random fashion across
the different regions within a factor of optical depths 1 $\sim$
5.

\subsection{Infall Candidates}

In this section we identify infall candidates based on the
spectral properties of the HCN line profiles together with those
of other lines (CS $J = 2-1$ and N$_{2}$H$^{+}$ $J = 1-0$ from
LMT99, and CS $J = 3-2$ and DCO$^{+}$ $J = 2-1$ from LMP04). The
spectral properties that we adopt for this purpose are (1) values
of $\delta$V for HCN $J = 1-0$  main component $F$=2$-$1 from this
study, for CS $J = 2-1$  and N$_{2}$H$^{+}$ from LMT99, and CS $J
= 3-2$ and DCO$^{+}$ from LMP04, and (2) peak intensity ratios of
the blue (T$_{b}$) to the red (T$_{r}$) component in double-peaked
spectra for HCN $F$=2$-$1 from this study, for CS $J = 2-1$ and
N$_{2}$H$^{+}$ from LMT99, and for CS $J = 3-2$ and DCO$^{+}$ from
LMP04.

The values of $\rm \delta V$ enable us to decide how much the
profiles are shifted with respect to the peak velocity of the
optically thin tracer. However, $\delta$V is sometimes affected by
the weak N$_{2}$H$^{+}$ $J = 1-0$ emission, even in the case that
the optically thick tracers have a double peaked infall asymmetry.
Peak intensity ratios are used to add more information in this
case. Moreover the peak intensity ratios select other cases where
the higher density tracers such as N$_{2}$H$^{+}$ and DCO$^{+}$
show a double peaked infall asymmetry. This is why we adopt
multiple spectral properties to choose the infall candidates.

The $\delta$V value is flagged with `1' for $\delta V \leq
-5\sigma_{\rm \delta V}$, `0' for $\mid \delta V \mid < 5
\sigma_{\rm \delta V}$, and `-1' for $\delta V \geq 5 \sigma_{\rm
\delta V}$. A good infall candidate should be given a number `1'.

In a similar way the intensity ratio is also flagged with `1' for
T$_{b}$/T$_{r}$ $>$ 1+ $\sigma_{\rm T_{b}/T_{r}}$, `0' for
$1-\sigma_{\rm T_{b}/T_{r}}$ $<$ T$_{b}$/T$_{r}$ $<$
$1+\sigma_{\rm T_{b}/T_{r}}$, and `-1' for T$_{b}$/T$_{r}$ $<$ $1-
\sigma_{\rm T_{b}/T_{r}}$, where the values of T$_{b}$ and T$_{r}$
were obtained from Gaussian fits of blue and red components of the
profile. Again, a dense core flagged with 1 should be considered
to be an infall candidate. The above criteria used for selecting
infall candidates are similar to those used by LMT99 and LMP04,
but have more data from this HCN study.

Table 2 summarizes ``color codes" of each source coupled with
above two spectral properties. These values can be used as
quantitative indicators for the selection of infall candidates
more accurately. Each source can be given the total of 9 color codes
although most of color codes for property (2) of DCO$^{+}$ and
N$_{2}$H$^{+}$ remain unassigned because these lines are usually
Gaussian except for several cores, such as L1521F, L1544, L492,
L694-2 and L1197. In the last column we list the total color codes
which are the algebraic sum of all color codes. We defined
``Probable'' infall candidate which has a total color codes
between 3 $\sim$ 4. Such candidates are B18-3, TMC2, TMC1, L1522, L1696A,
L1696B, L1689B, L234E-S, L63 and L1197. L183 was classified as a
possible infall candidate in LMP04 and a strong infall candidate
in LMT99. But it is not a strong infall candidate in our study
because it shows minus signs of asymmetry from both spectral
properties in the HCN. On the other hand, L63 and L1197 were not
grouped as infall candidate in LMP04 and LMT99, but they are
probable candidates from both spectral properties in this paper.
We classified 7 strong infall candidates which have a total color
code $ > $ 4: L1355, L1498, L1521F, L1544, L492, L694-2, and
L1155C1. All the strong infall candidates have been classified as
previously in LMP04, LMT99.

\section{DISCUSSION}

\subsection{Infall Structure}
As noted in the previous section, the HCN lines seem to be the
best tracer for the study of infall kinematics in the core not
only because they show infall asymmetry with clear dip more often
than other tracers, but also because they are composed of three
hyperfine lines with different opacities. These features enable us
to trace different layers of the core and makes it useful to study
infall structure along the line of sight of the core. Probably the
hyperfine line $F$=2$-$1 with the highest opacity would trace the
far side from the center of the core and the other hyperfine line
$F$=0$-$1 with the least opacity would trace the closer side from
the center of the core. So, if the hyperfine lines show the infall
asymmetry, then we may be able to see the variation of infall
velocities with the distance from the center by {\bf analyzing}
the HCN hyperfine lines. For this purpose, we choose the infall
candidates showing infall asymmetry in at least two hyperfine
lines and derive peak intensity ratios of the blue (T$_{b}$) to
the red (T$_{r}$) components in HCN hyperfine lines of the infall
candidates. Suppose that the excitation temperature increases
toward the core, infall speed of the gaseous material is small
(less than the velocity dispersion of the gas in the core), and
the central dip in the infall asymmetry profile is formed by the
absorption of the emission lines from the approaching warm part
near the central part of the core by the cool envelope receding
from an observer. Then the ratio of blue to red components should
be proportional to the gaseous infall speed. So with the ratios of
the blue to the red components, it may be possible to deduce
qualitative features of the infall speed in the core, although it
may not be possible to give any explicit form of the variational
infall speed with radius. Fig. 6 plots the intensity ratios of the
blue to the red components in HCN for each infall candidates,
showing that four sources such as L63, L492, L694-2, and L1197 out
of 12 infall candidates have larger ratio in the hyperfine line
with the lower opacity. If the excitation temperature increases
toward the center, these four cores may have increasing infall
speeds toward center of the cores.

However, the infall structure of the core and observational data
are not so simple in practice. LMP04 have shown from their CS
surveys that CS $J = 3-2 $ usually traces inner denser region in
faster inward motions than CS $J = 2-1$, indicating that infall
structures are developed in several cores (L1544, L1521F, L1445,
L183, L158, L1689B, and L694-2). Williams, Lee, \& Myers (2006)
also find faster inward motions in the inner denser region for
L1544 and L694-2 using their N$_{2}$H$^{+}$$ J = 1-0 $ data.
However, our HCN hyperfine lines have revealed such a structure
toward the L694-2 only.

Because the optical depths and excitation conditions for the
tracers are different among the sources, different tracers would
trace different regions of the cores. Therefore it would be
natural not to expect the detection of the same features of inward
motions in every core with different tracers.

Ambipolar Diffusion Models (e.g., Ciolek \& Basu 2000) usually
predict increasing infall speed toward the center up to certain
radius beyond which the infall speed decreases. The turning point
and the magnitude of the infall speed are dependent upon the
evolutionary stage of the core. Because the spectral line profile
always carries the average information through the line of sight,
we may be watching such a complication.

A detailed Monte Carlo radiative transfer model may help to deduce
the infall structure of the core from the complicated line
profiles. For example, Keto et al. (2004) made an effort to derive
the information of the density, velocity, and molecular abundance
of three starless cores L1544, L1489 and L1517B. A Monte Carlo
radiative transfer calculation by Lee et al. (2007) was
able to fit the infall profiles of the HCN hyperfine lines for
L694-2 and L1197 assuming approximate Bonner-Ebert density
distribution and $\Lambda$ shaped infall velocity structure
(increasing inward infall velocity to a point somewhere
between the center and the envelope of the core from where the
infall velocity starts to decrease toward the center of the core).

The mapping of the cores with HCN hyperfine lines will also help
us to understand spatial variation of the infall speed of the
cores to the direction orthogonal to the line of sight of the
core. Thus combining the mapping data and the detailed Monte Carlo
radiative transfer model would be very useful for better
understanding of the infall structure of the cores. The study in
this direction is underway.

\subsection{Less depletion of HCN and its effect}

Whether the HCN depletes out in the central region of the core is
an important factor in studying the inward motions in the cores.
We measure the line intensities of the HCN $F$=0$-$1 components
with the least optical depths among three hyperfine components for
36 sources showing the least self-absorbed features, to compare
with those of N$_{2}$H$^{+}$ which hardly suffers from the
molecular depletion in most cores and generally well traces
the density distribution of the cores (Tafalla et al. 2002; Bergin
\& Langer 1997). Figure 7 shows the correlation between the
integrated intensities of each HCN hyperfine line and
N$_{2}$H$^{+}$ line.
The intensity of N$_{2}$H$^{+}$ line is the total integrated intensity 
of its seven hyperfine components which is adopted from LMT99.
The integrated intensity of each HCN hyperfine line lines is simply  
a summation of the channel intensity multiplied by 
the channel width over each hyperfine line profile.

 It appears that HCN $F$=0$-$1 integrated intensities
have a very good correlation (correlation coefficient \emph{r} =
0.82) with those of N$_{2}$H$^{+}$. We believe that a slight
deviation from a tight relation between the intensities of two
lines is partly due to a rather large optical depth of the HCN
$F$=0$-$1 component. Our $\rm H^{13}CN$ observations enable us to
estimate its optical depths of $1\sim 3$ for 8 sources. The worse
correlations of the intensities of the other HCN components with
higher optical depths (Fig. 7-b and c) support this
possible effect of high optical depth on the HCN lines.

Our follow-up mapping observations in HCN also support the idea
that the HCN is less affected from the molecular depletion in the
central region of the core, from the fact that HCN $F$=0$-$1 has
similar spatial distribution to that of N$_{2}$H$^{+}$ (Sohn et
al. 2007 in prep.). Altogether the HCN suffers from less depletion
in the central region and therefore it is thought to trace inner
regions of the cores better than the CS. 
This may result in the
higher occurrence of infall motions of gaseous molecules and more
negative $\delta V$ for HCN than that of CS.

As another evidence for the HCN to be a good tracer of the central region 
of the core, one can suggest some possible correlation between the degree of the self-absorption 
in the HCN line and the optical depth of the $\rm N_2H^+$ line. 
We tested the relation in 17 sources showing a clear self-absorption in HCN(1-0) hyperfine  
by measuring $\rm (T_{peak} -T_{dip}) / T_{peak}$ of the HCN profiles and the ratio of the line widths of 
the emission profile and the absorbed feature, as indicators of the amount of the self-absorption.
Here $\rm T_{peak}$ is the peak temperature 
of the observed HCN spectrum and $\rm T_{dip}$ is the intensity of the self-absorbed dip.  
These quantities were compared with the tau values of $\rm N_2H^+$ line taken from LMT09,
but there were found to be no clear correlation between them. 
In fact the degree of self-absorption itself in a molecular line toward 
the cores can be affected not only by the optical depth of the tracer,
but also by other properties such excitation conditions and kinematics in the cores 
like infall motions. 
In addition, once a spectrum is self-absorbed, it is practically difficult to measure  
physical quantities of the real dip and the unabsorbed profile, without employing a proper model 
dealing with radiation processes in the core which 
is actually beyond the scope of this paper. 
We suggest that at least the property of the less depletion of the HCN 
in the inner region of the core
can be a preferable evidence for the HCN to be a useful tracer of the central regions of the core.  

\section{SUMMARY}
We performed a systematic pointing survey of starless cores in HCN
and/or $\rm H^{13}CN$ $ J = 1-0 $ hyperfine transitions to study
inward motions in starless cores. Sixty-four and ten of 85 targets
were detected in HCN and $\rm H^{13}CN$ lines, respectively. Main
results obtained from the survey are as follows;

1. Most of the HCN spectra detected toward the starless cores are
found to exhibit asymmetric features such as the blue asymmetry
(43 \%), the red asymmetry (20 \%), and the mixture of blue and
red asymmetry in the hyperfine lines (33 \%).

2. The HCN lines  show the highest occurrence of double peaked
features (for more than 30 \% of the detected cores), more
asymmetric feature, and less depletion toward the central region
of the cores than other infall tracers. Especially three HCN
hyperfine lines with different optical depths are useful to study
kinematics of the different regions of the cores. Altogether HCN
lines are found to be one of the best infall tracers.

3. The statistical distribution of the velocity shift $\delta V$
of the HCN profiles with respect to the systemic velocity of the
optically thin tracer is found to be significantly shifted to the
blue side, implying the predominance of the inward motions in the
starless cores. The degree of the skewness in the distribution is
greater than that of any other infall tracers, indicating that the
HCN traces more frequently inward motions. The $\delta V$
distributions of HCN hyperfine lines are similar within the error.
Moreover the $\delta V$ values of HCN hyperfine lines for each
source are also similar. These may mean that most of starless
cores have inward motions in a systematic rather than random
fashion across the layers of the cores.

4. This survey enables us to identify infall candidates for
further study in the future, using all available ``color codes''
such as the velocity shift $\delta V$ and the blue to red peak
intensity ratio of double peaked profiles for HCN $ J = 1-0 $, CS
$ J = 2-1 $ , $ J= 3-2$, DCO$^+$ $ J = 2-1 $, and $\rm N_2H^+$ $ J
= 1-0 $. We identify 7 strong infall candidates: L1355, L1498,
L1521F, L1544, L492, L694-2, and L1155C1, and 10 probable infall
candidates: B18-3, TMC2, TMC1, L1552, L1696A, L1696B, L1689B, L234E-S,
L63, and L1197.

5. We find four infall candidates such as L63, L492, L694-2, and
L1197 with the higher blue to red ratio in the hyperfine line of
the lower opacity, suggesting that they may be in a stage of
development of infall structure of increasing infall speeds
inside.

\acknowledgments

This research was supported by the KOSEF grant 
No. R14-2002-058-01000-0 of the Korea Science and Engineering
Foundation and in part by KOSEF grant No. KOSEF
R01-2003-000-10513-0 of the Korea Science and Engineering
Foundation. The authors express special thanks to the staffs of
Taeduk 14m observatory for their dedicated support.




\begin{figure}[btp]
\epsscale{0.8} \plotone{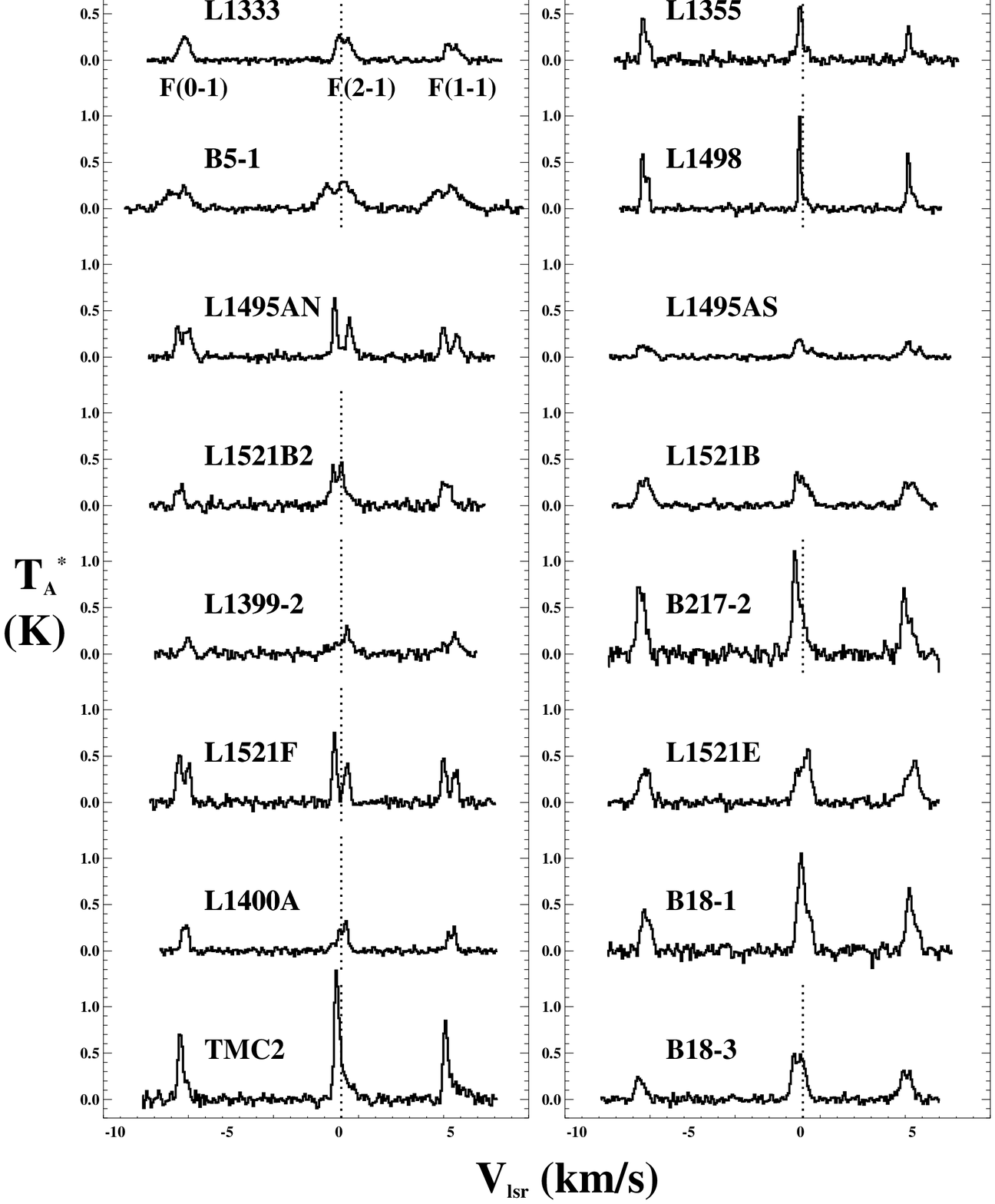} \caption{HCN $J= 1-0$ profiles
for 64 sources with strong detection. All profiles are shifted so
that the $\rm V_{N_2H^+}$ or Gaussian fit velocity of HCN
$F$=2$-$1 component, in case of no existence of $\rm V_{N_2H^+}$,
is zero. \label{fig1}}
\end{figure}

\begin{figure}[btp]
\epsscale{0.9}
\plotone{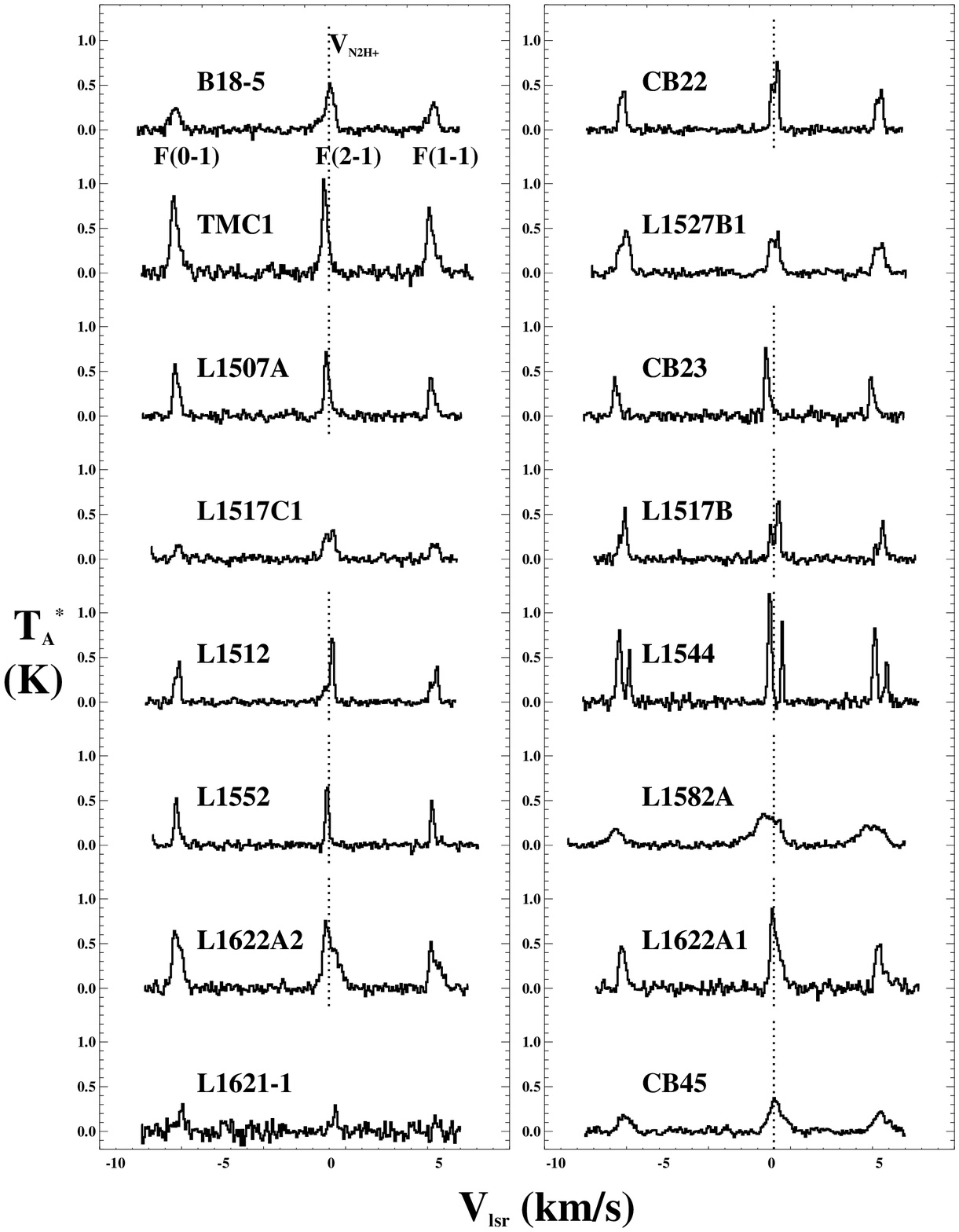}
\end{figure}

\begin{figure}[btp]
\epsscale{0.9}
\plotone{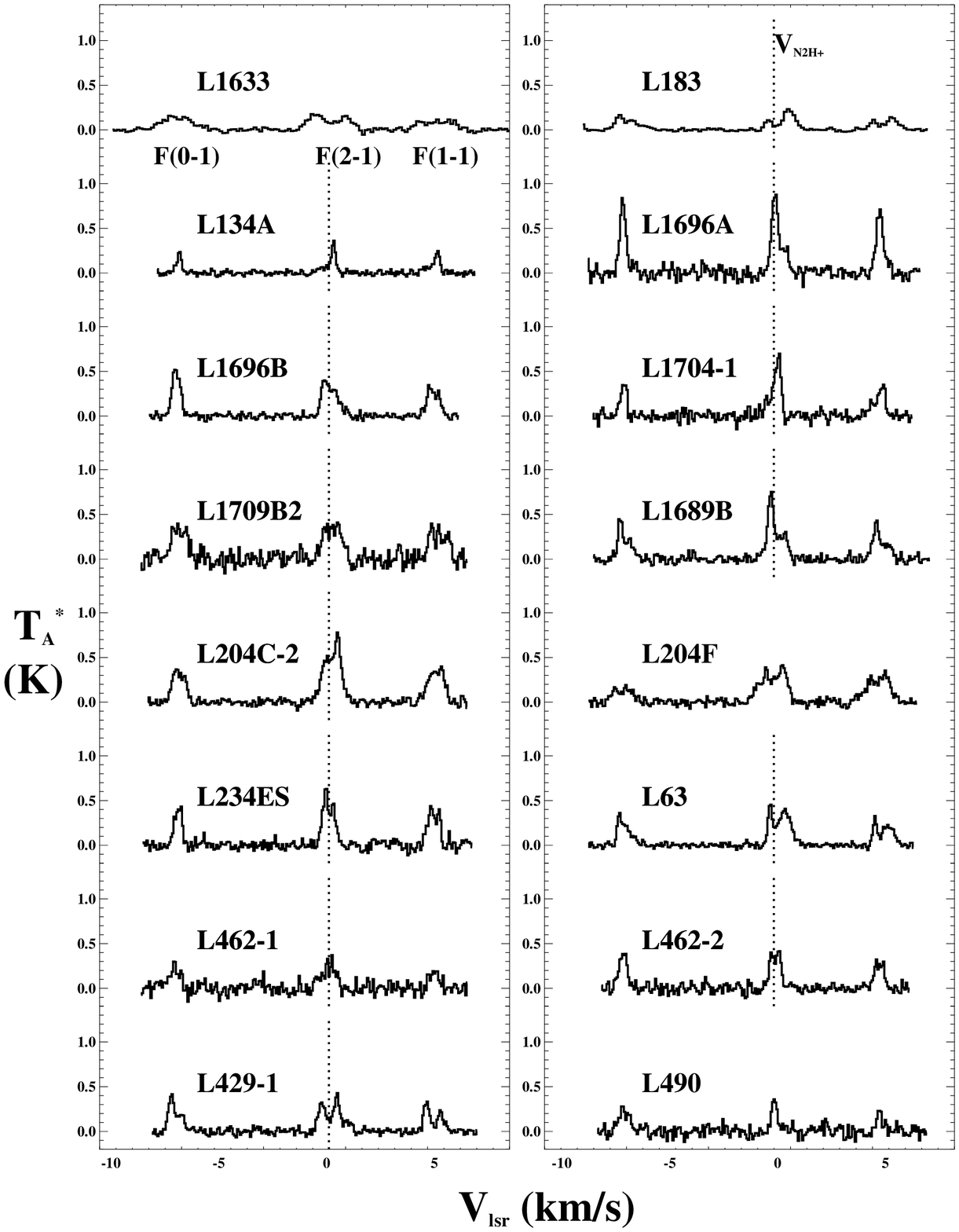}
\end{figure}

\begin{figure}[btp]
\epsscale{0.9}
\plotone{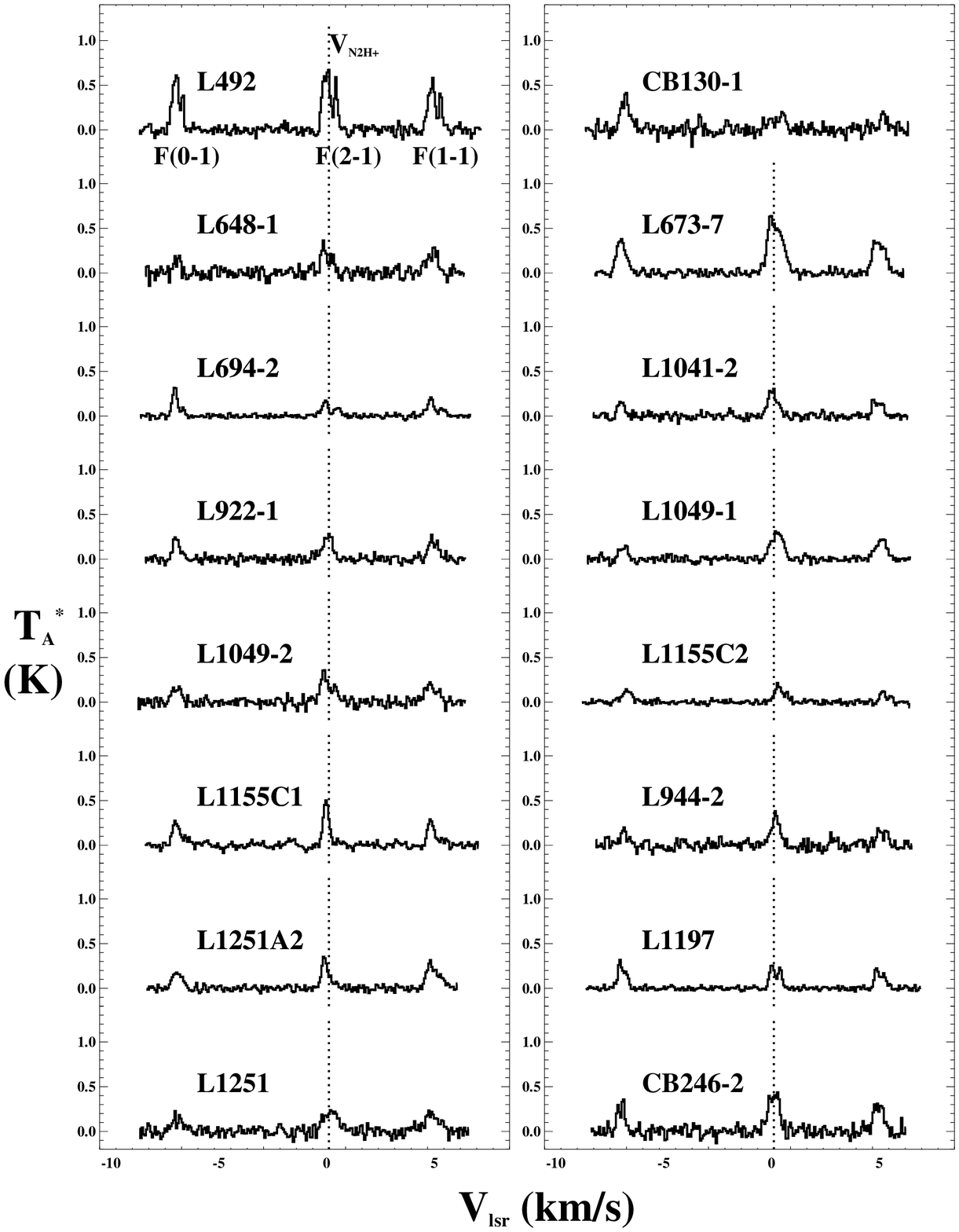}
\end{figure}

\begin{figure}[btp]
\plotone{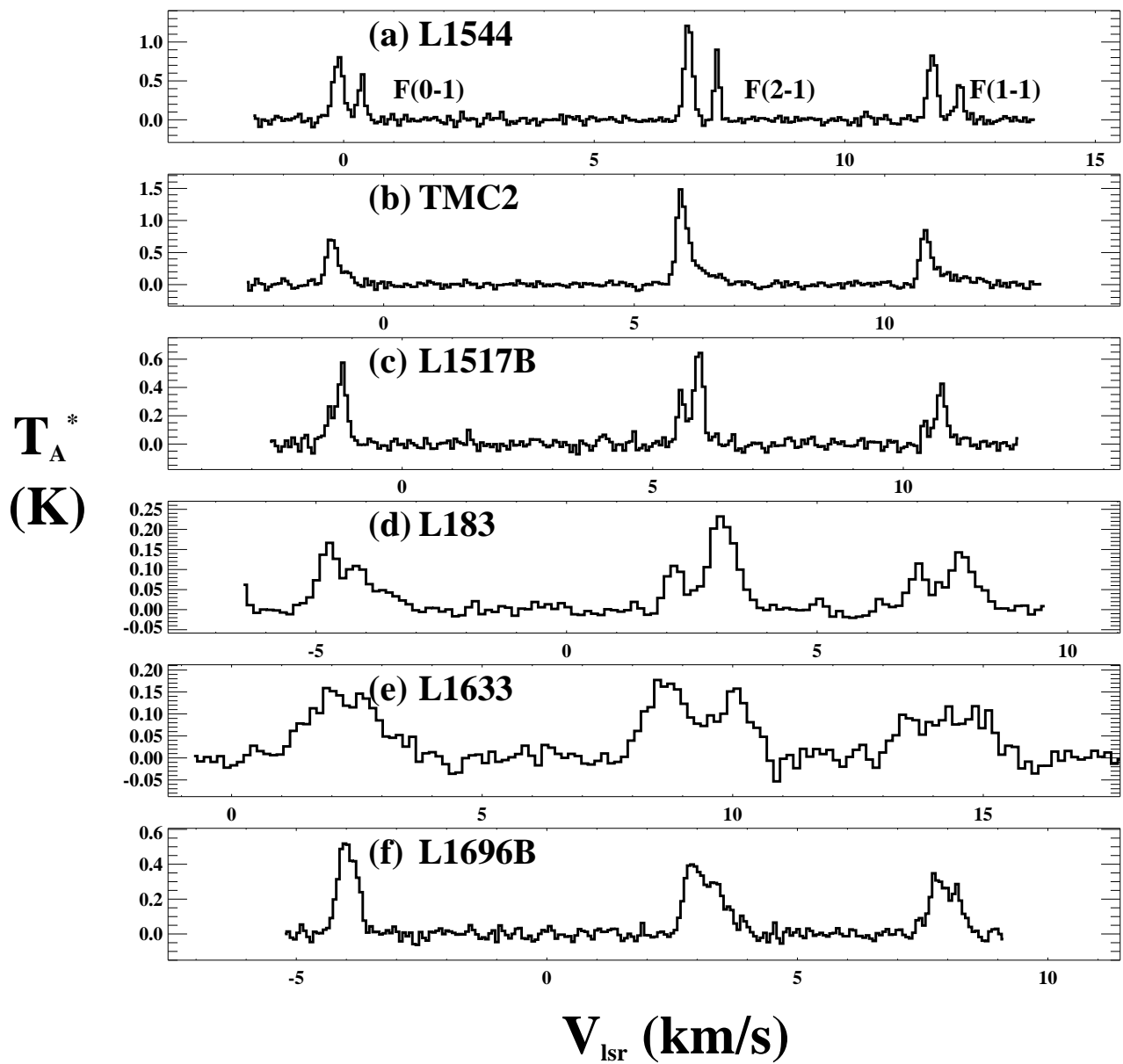} \caption{Six prototypical HCN $J=1-0$ spectra. (a)
All three components  have two peaks, with the blue peak brighter
than the red. (b) One or two of hyperfines has two peaks or single
peak of the infall asymmetry. (c) At least one of hyperfines has
red asymmetry. (d) Red asymmetry and blue asymmetry are mixed in
three hyperfines. (e) The line profiles are broad. (f) Hyperfine
intensity ratios are anomalous. \label{fig3}}
\end{figure}

\clearpage
\begin{figure}[btp]
\plotone{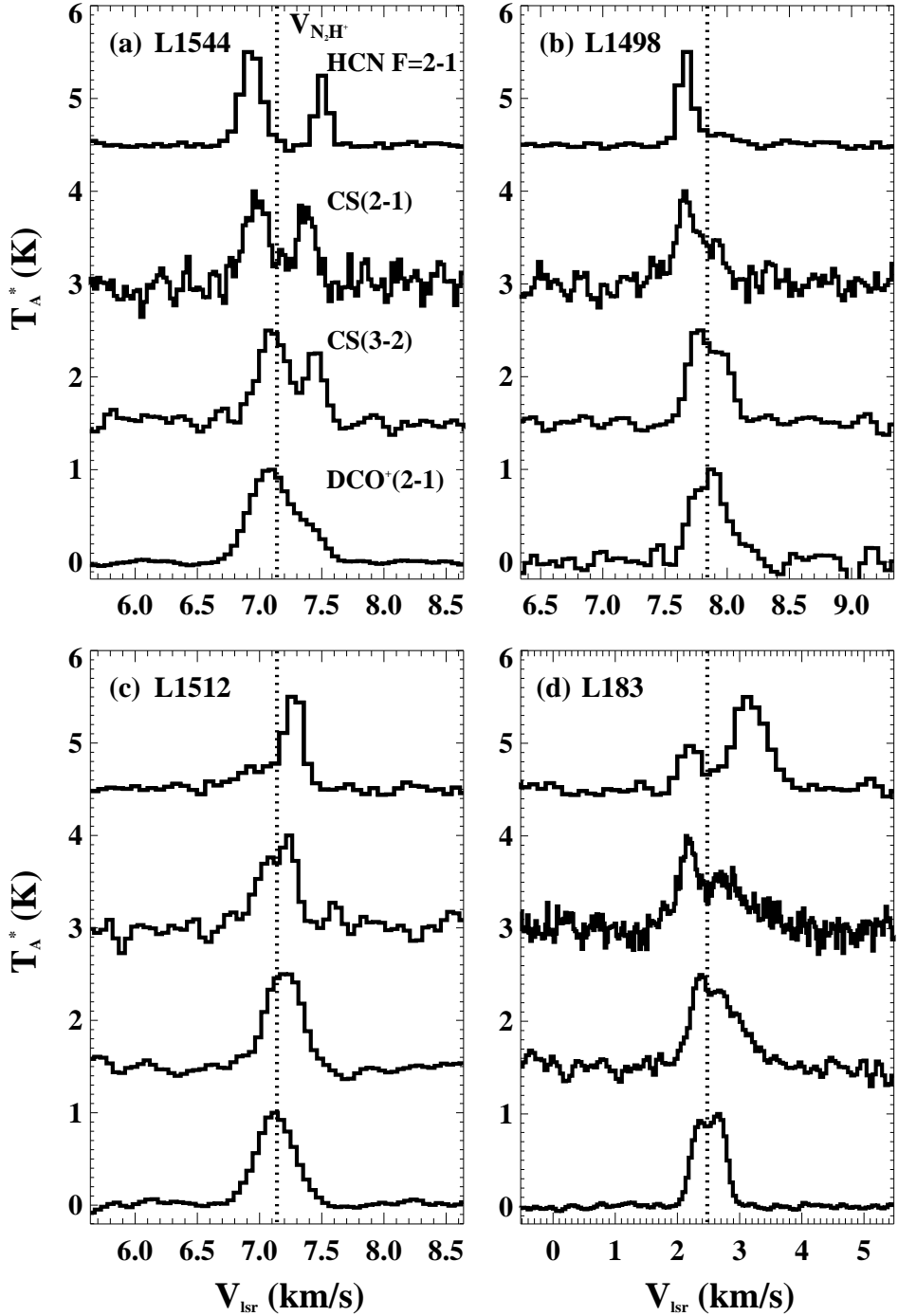} \caption{HCN profiles compared with those of other
infall tracers. (a) The HCN profile is usually the most opaque
among other infall tracers. (b) Either the blue profiles or (c)
the red profiles of the HCN show more asymmetric features. (d) The
HCN profiles sometimes show totally revered asymmetry to those of
other line profiles. \label{fig4}}
\end{figure}

\begin{figure}[btp]
\epsscale{0.7} \plotone{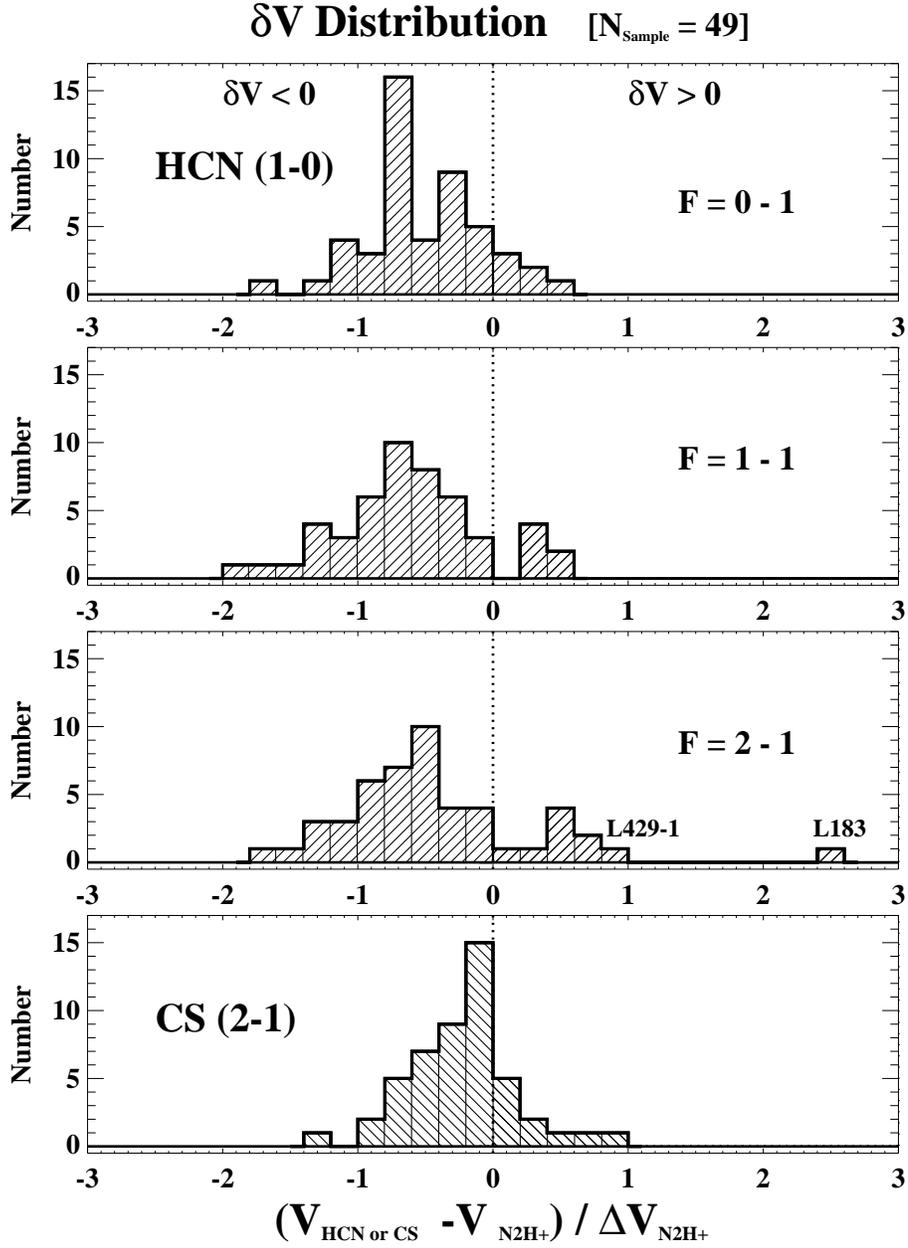} \caption{Histograms of the
normalized velocity difference ($\delta V_{\rm HCN \ \it J \rm
=1-0}$) between $V_{\rm HCN(F=i-j)}$ and $V_{\rm N_{2}H^{+} \ \it
J \rm =1-0}$ for a sample of 49 starless cores. The distribution
of $\rm \delta V_{CS \ \it J \rm =2-1}$ from LMT99 is also given
after adopting the frequency set for CS and $\rm N_2H^+$
determined from LMT01. The $\rm \delta V$ distribution of each HCN
hyperfine transition appears to be similar, but more skewed to the
blue than that of $\rm \delta V_{CS \ \it J \rm =2-1}$.
\label{fig4}}
\end{figure}

\begin{figure}[btp]
\epsscale{1.0} \plotone{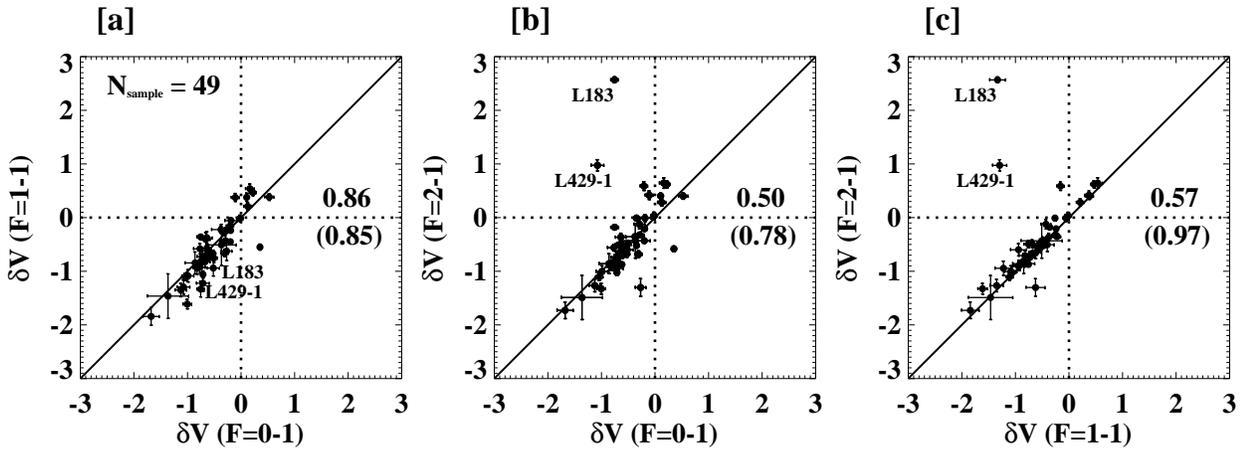} \caption{Correlations among
$\delta$Vs of each HCN hyperfine line. The solid lines indicate
the line for perfect correlation between different hyperfines. The
numeric number at right of each panel is a correlation coefficient
of the fit. Parenthesized numbers are the correlation coefficients
without L183 and L429-1. \label{fig5}}
\end{figure}

\clearpage
\begin{figure}[btp]
\epsscale{0.9} \plotone{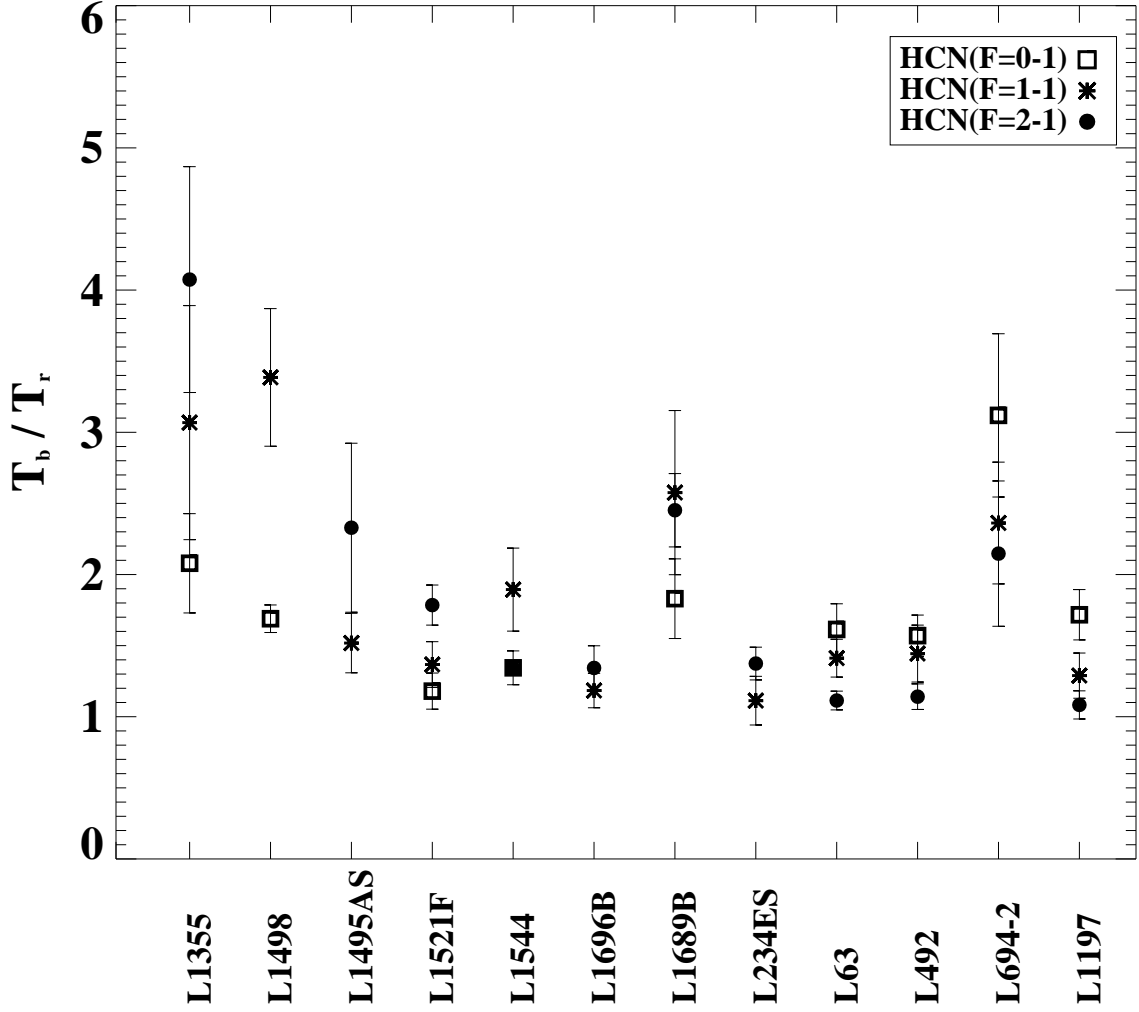} \caption{Intensity ratios of the
blue (T$_{b}$) to the red (T$_{r}$) components in HCN hyperfine
lines. Note that four infall candidates, L63, L492, L694-2, and
L1197 show the larger value for T$_{b}$/T$_{r}$ in the hyperfine
line with the lower opacity. \label{fig6}}
\end{figure}

\clearpage
\begin{figure}[btp]
\epsscale{0.3} \plotone{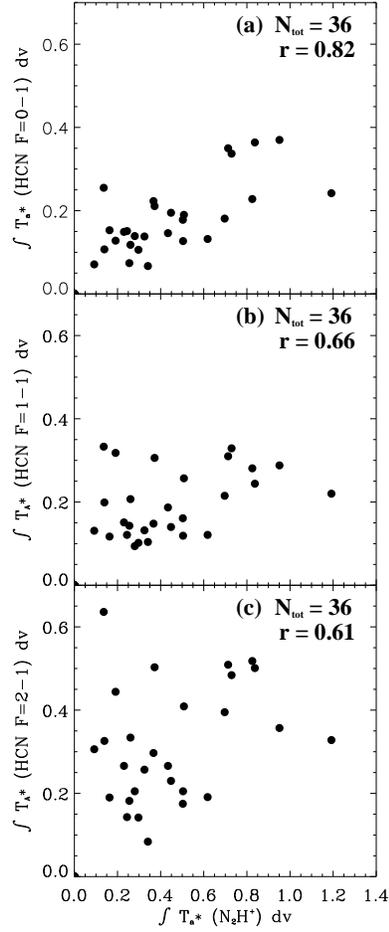} \caption{The comparison between the
integrated intensities of each HCN $J=1-0$ hyperfine and
N$_{2}$H$^{+}$ $J=1-0$. The diagrams show that the intensity of
the HCN ($F$=0$-$1) has better correlation (correlation
coefficient \emph{r} = 0.82) with the intensity of N$_{2}$H$^{+}$
than those of other HCN hyperfine lines. \label{fig7}}
\end{figure}

\end{document}


\begin{landscape}
\small
\begin{deluxetable}{lrrrcrrrrrrrl} 
\setlength{\tabcolsep}{0.8mm} \tabletypesize{\scriptsize}
\tablecaption{The line parameters of the starless cores observed
in HCN and $\rm H^{13}CN$ $J=1-0$ lines}
\tablewidth{0pt}
\tablehead{
       \colhead{ } & \colhead{}  & \colhead{} &\colhead{}
&\colhead{}&\multicolumn{3}{c}{V$_{HCN}$ (km s$^{-1}$) }
     & \colhead{ } & \colhead{ } & \multicolumn{3}{c}{$\delta$V$_{HCN}$ } \\
        \colhead{ } & \colhead{R.A.(2000)}  &\colhead{Dec.(2000)} &
\colhead{T$_A^*$} &
       \colhead{T$_A^*$(H$^{13}$CN)}
&\multicolumn{3}{l}{-----------------------------------------------}
       & \colhead{V $_{N_{2}H^{+}}$ }
       & \colhead{$\Delta V_{N_{2}H^{+}}$ }
&\multicolumn{3}{l}{-----------------------------------------------} \\
        \colhead{Source}         & \colhead{$^h~~^m~~^s~$}
&\colhead{~~~~$^{\circ}~~~'~~~''$} &
        \colhead{F =2-1} & \colhead{F=2-1} & \colhead{[F= 0-1}       &
\colhead{F =1-1}         &\colhead{F =2-1]} &
        \colhead{(km s$^{-1}$)} & \colhead{(km s$^{-1}$)} & \colhead{[F=
0-1}       & \colhead{F =1-1}         &\colhead{F =2-1]}}
\startdata
  L1333 & 02~26~13.7 & 75~27~03 & 0.27    &  \nodata& 3.72$\pm$0.02 & 3.63
$\pm$ 0.01 &  3.64 $\pm$ 0.01 &  3.80$\pm$0.01
  & 0.38 $\pm$ 0.01 &  -0.200 $\pm$ 0.03 &  -0.450 $\pm$ 0.04 &  -0.432
$\pm$0.04\\
  L1355 & 02~53~12.2 &68~55~52 &0.58    &  \nodata&-3.92 $\pm$ 0.01 & -3.96
$\pm$ 0.01 & -3.93 $\pm$ 0.01 & -3.75$\pm$0.01
  & 0.26 $\pm$ 0.03 &  -0.669 $\pm$ 0.09 &  -0.819 $\pm$ 0.11 &  -0.708
$\pm$0.09\\
  B5-1& 03~47~36.1 &32~55~58 &0.28    &  \nodata & 9.89 $\pm$ 0.02 &  9.90
$\pm$ 0.02 & 9.94 $\pm$ 0.04 &  9.91$\pm$0.06
  & 1.03 $\pm$ 0.11 &  -0.018 $\pm$ 0.06 &  -0.015 $\pm$ 0.06 &  0.031 $\pm$
0.07\\
  B5-2 & 03~48~25.3 &32~54~17 & $<$ 0.21 &  \nodata &  \nodata        &
\nodata        &   \nodata        &   \nodata
  &  \nodata  &  \nodata  &   \nodata  &  \nodata \\
  L1498 & 04~10~51.5 &25~09~58 &1.00    & 0.20 & 7.61 $\pm$ 0.00 &  7.59
$\pm$ 0.00 &  7.61 $\pm$ 0.00    &  7.82$\pm$0.01
  & 0.21 $\pm$ 0.02 & -0.990 $\pm$ 0.11 &  -1.086 $\pm$ 0.11 &  -1.005
$\pm$0.11\\
  L1495AN &04~18~31.8 &28~27~30 &0.65   &  \nodata &  7.03 $\pm$ 0.01 &
7.00 $\pm$ 0.01 &  7.01 $\pm$ 0.02    &  7.30$\pm$0.01
  & 0.36 $\pm$ 0.03 & -0.749 $\pm$ 0.07 &  -0.848 $\pm$ 0.10 &  -0.810
$\pm$0.08\\
  L1495AS&04~18~42.0 &28~23~50 &0.20   &  \nodata &  7.08 $\pm$ 0.03 &  7.14
$\pm$ 0.01 &  7.14 $\pm$ 0.02    &  7.31$\pm$0.01
  & 0.30 $\pm$ 0.02 & -0.767 $\pm$ 0.12 &  -0.581 $\pm$ 0.09 &  -0.561
$\pm$0.06\\
  L1495-2&04~17~57.7 &28~37~58 &$<$ 0.25 &  \nodata &  \nodata        &
\nodata        &   \nodata        &   \nodata
  &  \nodata  &  \nodata  &   \nodata  &  \nodata \\
  L1495B &04~18~05.1 &28~22~22 &$<$ 0.26 &  \nodata &  \nodata        &
\nodata        &   \nodata        &   \nodata
  &  \nodata  &  \nodata  &   \nodata  &  \nodata \\
  L1521B-2\tablenotemark{a}&04~23~37.0 &26~40~06 &0.43  &  \nodata  & 6.29 $\pm$ 0.01 &  6.04
$\pm$ 0.03 &  6.02 $\pm$ 0.01 & 6.43$\pm$0.02
  & 0.19 $\pm$ 0.07 &  -0.758 $\pm$ 0.31 &  -2.053 $\pm$ 0.79 &
-2.184$\pm$0.81\nl
  L1521B&04~24~12.7 &26~36~53 &0.36    &  \nodata &  \nodata        &
\nodata        &      \nodata     &   \nodata
  &  \nodata  &  \nodata  &   \nodata  &  \nodata \\
  L1399-2&04~24~58.8 &54~59~40 &0.28   &  \nodata & \nodata &  \nodata
&\nodata   & 3.39$\pm$0.01
  & 0.45 $\pm$ 0.03 &  \nodata &  \nodata  & \nodata \\
  L1399-3&04~25~12.2 &54~56~50 &$<$ 0.15 &  \nodata &  \nodata        &
\nodata        &   \nodata        &   \nodata
  &  \nodata  &  \nodata  &   \nodata  &  \nodata \\
  B217-2&04~28~08.6 &26~20~53 &1.09    &  \nodata& 6.45 $\pm$ 0.01 &  6.43
$\pm$ 0.01 & 6.43 $\pm$ 0.01   & 6.84$\pm$0.01
  & 0.37 $\pm$ 0.02 &  -1.041 $\pm$ 0.07 &  -1.108 $\pm$ 0.07 &
-1.105$\pm$0.07\\
  L1521F&04~28~39.8 &26~51~35 &0.73    & $<$ 0.22 & 6.24 $\pm$ 0.01 &  6.12
$\pm$ 0.02 & 6.13 $\pm$ 0.01  & 6.49$\pm$0.00
  & 0.35 $\pm$ 0.01 &  -0.711 $\pm$ 0.03 &  -1.060 $\pm$ 0.07 &
-1.023$\pm$0.03\\
  L1521E&04~29~13.6 &26~14~19 &0.54    &  \nodata &  \nodata        &
\nodata        &  \nodata         &   \nodata
  &  \nodata  &  \nodata  &   \nodata  &  \nodata \\
  L1400A &04~30~56.8 &54~52~36 &0.31    &  \nodata & 3.33 $\pm$ 0.01 &  3.41
$\pm$ 0.01 & 3.44 $\pm$ 0.01  & 3.29$\pm$0.01
  & 0.23 $\pm$ 0.03 &  0.165 $\pm$ 0.08 &  0.539 $\pm$ 0.09 &
0.643$\pm$0.10\\
  B18-1 &04~31~57.7 &24~32~30 &1.03    &  \nodata &  \nodata        &
\nodata        &   \nodata        &   \nodata
  &  \nodata  &  \nodata  &   \nodata  &  \nodata \\
  L1445&04~32~07.8 &46~37~23 &$<$ 0.25 &  \nodata&  \nodata        &
\nodata        &   \nodata        &   \nodata
  &  \nodata  &  \nodata  &   \nodata  &  \nodata\\
  B18-2&04~32~36.7 &24~52~05 &$<$ 0.27 &  \nodata &  \nodata        &
\nodata        &   \nodata        &   \nodata
  &  \nodata  &  \nodata  &   \nodata  &  \nodata \\
  TMC2&04~32~49.0 &24~25~12 &1.50    &  0.17& 5.96 $\pm$ 0.01 &  5.93 $\pm$
0.03 & 5.94 $\pm$ 0.00  & 6.21$\pm$0.01
  & 0.30 $\pm$ 0.02 &  -0.820 $\pm$ 0.07 &  -0.940 $\pm$ 0.12 &
-0.910$\pm$0.07\\
  B18-3&04~34~06.1 &24~09~45 &0.49   &  \nodata & 6.53 $\pm$ 0.03 &  6.50
$\pm$ 0.01 & 6.49 $\pm$ 0.01  & 6.94$\pm$0.03
  & 0.30 $\pm$ 0.08 &  -1.363 $\pm$ 0.38 &  -1.463 $\pm$ 0.41 &
-1.490$\pm$0.41\\
  B18-5&04~35~51.3 &24~09~21 &0.50    &  \nodata& 6.00 $\pm$ 0.04 &  6.10
$\pm$ 0.02 & 6.11 $\pm$ 0.02  & 6.24$\pm$0.02
  & 0.37 $\pm$ 0.04 &  -0.638 $\pm$ 0.10 &  -0.378 $\pm$ 0.12 &
-0.365$\pm$0.07\\
  CB22&04~40~39.9 &29~52~59 &0.77    &  \nodata& 5.86 $\pm$ 0.01 &  5.89
$\pm$ 0.01 & 5.97 $\pm$ 0.01  & 5.97$\pm$0.01
  & 0.31 $\pm$ 0.03 & -0.345 $\pm$ 0.05 &  -0.261 $\pm$ 0.05 &  -0.010
$\pm$0.05\\
  TMC1&04~41~33.0 &25~44~44 &1.08    &  0.23 & 5.62 $\pm$ 0.01 &  5.56 $\pm$
0.01 & 5.57 $\pm$ 0.01  & 5.86$\pm$0.01
  & 0.43 $\pm$ 0.02 &  -0.560 $\pm$ 0.04 & -0.693 $\pm$ 0.06 &
-0.670$\pm$0.04\\
  L1527B-1&04~41~33.0 &25~46~24 &0.57 &  \nodata&  \nodata        &
\nodata        &   \nodata        &   \nodata
  &  \nodata  &  \nodata  &   \nodata  &  \nodata\\
  L1507A&04~42~38.6 &29~43~45 &0.74   &  \nodata & 6.03 $\pm$ 0.01 &  6.00
$\pm$ 0.01 & 6.00 $\pm$ 0.01  & 6.20$\pm$0.01
  & 0.22 $\pm$ 0.02 &  -0.768 $\pm$ 0.09 &  -0.923 $\pm$ 0.10 &
-0.891$\pm$0.10\\
  CB23&04~43~31.5 &29~39~11 &0.78    &  \nodata& 5.64 $\pm$ 0.01 &  5.60
$\pm$ 0.01 & 5.63 $\pm$ 0.01  & 6.04$\pm$0.01
  & 0.24 $\pm$ 0.02 & -1.679 $\pm$ 0.15 &  -1.847 $\pm$ 0.16 &
-1.729$\pm$0.15\\
  L1517C-1&04~54~47.4 &30~35~38 &0.34   &  \nodata &  \nodata        &
\nodata        &  \nodata         &   \nodata
  &  \nodata  &  \nodata  &   \nodata  &  \nodata\\
  L1517B&04~55~18.8 &30~38~04 &0.67    &  \nodata& 5.83 $\pm$ 0.01 &  5.89
$\pm$ 0.01 & 5.93 $\pm$ 0.01  & 5.77$\pm$0.01
  & 0.25 $\pm$ 0.02 &  0.220 $\pm$ 0.07 &  0.468 $\pm$ 0.06 &
0.620$\pm$0.07\\
  L1512&05~04~09.7 &32~43~09 &0.73    &  \nodata& 7.01 $\pm$ 0.01 &  7.07
$\pm$ 0.01 & 7.09 $\pm$ 0.00  & 7.12$\pm$0
  & 0.15 $\pm$ 0.01 & -0.753 $\pm$ 0.07 &  -0.360 $\pm$ 0.05 &
-0.180$\pm$0.02\\
  L1544&05~04~14.9 &25~11~08 &1.27   & 0.23 & 6.91 $\pm$ 0.01 &  6.89 $\pm$
0.01 & 6.88 $\pm$ 0.00  & 7.12$\pm$0.01
  & 0.34 $\pm$ 0.01 &  -0.606 $\pm$ 0.04 & -0.685 $\pm$ 0.04 &
-0.700$\pm$0.04\\
  L1523&05~06~16.4 &31~41~30 &$<$ 0.24&  \nodata&  \nodata        &
\nodata        &   \nodata        &   \nodata
  &  \nodata  &  \nodata  &   \nodata  &  \nodata \\
  L1552&05~17~37.4 &26~05~28 &0.69    &  \nodata& 7.53 $\pm$ 0.01 &  7.47
$\pm$ 0.00 & 7.53 $\pm$ 0.00  & 7.64$\pm$0.01
  & 0.22 $\pm$ 0.02 &  -0.509 $\pm$ 0.07 & -0.755 $\pm$ 0.08 &
-0.495$\pm$0.07\\
  L1582A&05~32~03.4 &12~31~05 &0.34   &  \nodata & 9.83 $\pm$ 0.04 &  9.56
$\pm$ 0.02 & 9.71 $\pm$ 0.03  & 10.21$\pm$0.03
  & 0.53 $\pm$ 0.06 &  -0.713 $\pm$ 0.12 & -1.223 $\pm$ 0.16 &
-0.945$\pm$0.13\\
  L1622A2&05~54~38.8 &01~53~44 &0.76  &  0.21  & 0.94 $\pm$ 0.01 &  0.90
$\pm$ 0.01 & 0.90 $\pm$ 0.01  &  1.10$\pm$0.01
  & 0.30 $\pm$ 0.01 &  -0.527 $\pm$ 0.05 & -0.667 $\pm$ 0.05 &
-0.677$\pm$0.06\\
  L1622A1&05~54~53.5 &01~57~24 &0.90  &  \nodata  & 1.03 $\pm$ 0.02 &  1.08
$\pm$ 0.01 & 1.03 $\pm$ 0.01  &  1.15$\pm$0.01
  & 0.33 $\pm$ 0.01 &  -0.373 $\pm$ 0.18 & -0.224 $\pm$ 0.07 &
-0.355$\pm$0.04\\
  L1621-1&05~55~59.4 &02~18~02 &0.29  &  \nodata  &  \nodata        &
\nodata        &  \nodata         &   \nodata
  &  \nodata  &  \nodata  &   \nodata  &  \nodata \\
  CB37&06~00~44.8 &31~39~29 &$<$ 0.33&  \nodata&  \nodata        &   \nodata
        &   \nodata        &   \nodata
  &  \nodata  &  \nodata  &   \nodata  &  \nodata \\
  L1574&06~08~05.0 &18~28~12 &$<$ 0.24 &  \nodata&  \nodata        &
\nodata        &   \nodata        &   \nodata
  &  \nodata  &  \nodata  &   \nodata  &  \nodata \\
  CB45&06~08~57.2 &17~50~15 &0.36  &  \nodata  & 0.19 $\pm$ 0.03 &  0.21
$\pm$ 0.02 & 0.17 $\pm$ 0.01  &  0.35$\pm$0.02
  & 0.56 $\pm$ 0.04 &  -0.286 $\pm$ 0.08 &  -0.246 $\pm$ 0.06 &
-0.323$\pm$0.05\\
  L1633&06~24~41.7 &03~22~11 &0.18  &  \nodata  &  \nodata        &
\nodata        &   \nodata        &   \nodata
  &  \nodata  &  \nodata  &   \nodata  &  \nodata \\
  L134A&15~53~33.1 &-04~35~26&0.36  &  \nodata  & 2.72 $\pm$ 0.01 &  2.82
$\pm$ 0.02 & 2.83 $\pm$ 0.01  & 2.68$\pm$0.01
  & 0.37 $\pm$ 0.02 &  0.105 $\pm$ 0.04 &  0.378 $\pm$ 0.08 &
0.403$\pm$0.05\\
  L183&15~54~06.6 &-02~52~23&0.23   &  \nodata & 2.28 $\pm$ 0.02 &  2.14
$\pm$ 0.04 & 3.08 $\pm$ 0.01  & 2.46$\pm$00.0
  & 0.24 $\pm$ 0.00 &  -0.754 $\pm$ 0.07 & -1.333 $\pm$ 0.15 &
2.567$\pm$0.05\\
  L1696A&16~28~31.4 &-24~19~08&0.89  &  \nodata  & 3.33 $\pm$ 0.02 &  3.36
$\pm$ 0.01 & 3.37 $\pm$ 0.01  & 3.37$\pm$0.01
  & 0.23 $\pm$ 0.02 & -0.187 $\pm$ 0.11 &  -0.057 $\pm$ 0.06 &
-0.009$\pm$0.07\\
  L1696B&16~28~59.4 &-24~20~43&0.41  &  \nodata  & 3.06 $\pm$ 0.01 &  2.92
$\pm$ 0.01 & 2.99 $\pm$ 0.02  & 3.29$\pm$0.01
  & 0.23 $\pm$ 0.01 &  -1.004 $\pm$ 0.08 &  -1.613 $\pm$ 0.09 &
-1.326$\pm$0.10\\
  L1704-1&16~30~50.7 &-23~42~14&0.72 &  \nodata   & 2.64 $\pm$ 0.01 &  2.76
$\pm$ 0.02 & 2.77 $\pm$ 0.01  & 2.67$\pm$0.01
  & 0.24 $\pm$ 0.03 &  -0.108 $\pm$ 0.07 &  0.375 $\pm$ 0.08 &
0.417$\pm$0.08\\
  L1709B-2&16~32~50.5 &-23~52~39&0.40 &  \nodata   & \nodata  &  \nodata  &
\nodata   & 2.56$\pm$0.01
  & 0.34 $\pm$ 0.02 & \nodata  & \nodata  &  \nodata \\
  L1689B&16~34~45.9 &-24~37~51&0.75  &  \nodata  & 3.30 $\pm$ 0.01 &  3.30
$\pm$ 0.02 & 3.30 $\pm$ 0.01  & 3.49$\pm$0.02
  & 0.22 $\pm$ 0.04 & -0.859 $\pm$ 0.18 &  -0.845 $\pm$ 0.21 &
-0.859$\pm$0.18\\
  L1782-2&16~42~32.8 &-19~43~37&$<$ 0.39 &  \nodata&  \nodata        &
\nodata        &   \nodata        &   \nodata
  &  \nodata  &  \nodata  &   \nodata  &  \nodata \\
  L158-2&16~46~52.2 &-13~58~45&$<$ 0.30 &  \nodata&  \nodata        &
\nodata        &  \nodata         &   \nodata
  &  \nodata  &  \nodata  &   \nodata  &  \nodata \\
  L204C-2&16~47~46.3 &-12~23~19&0.77  &  \nodata  & 4.15 $\pm$ 0.02 &  4.17
$\pm$ 0.02 & 4.50 $\pm$ 0.01  & 4.24$\pm$0.02
  & 0.45 $\pm$ 0.05 &  -0.207 $\pm$ 0.07 & -0.158 $\pm$ 0.07 &  0.587
$\pm$0.08\\
  L204F&16~47~48.5 &-11~56~06&0.40   &  \nodata &  \nodata        &
\nodata        &  \nodata         &   \nodata
  &  \nodata  &  \nodata  &   \nodata  &  \nodata \\
  L191-2&16~47~50.0 &-12~52~53&$<$ 0.30&  \nodata&  \nodata        &
\nodata        &  \nodata & \nodata
  &  \nodata  &  \nodata  &   \nodata  &  \nodata \\
  L234E-C&16~48~05.5 &-10~52~17&$<$0.24&  \nodata&  \nodata        &
\nodata        &   \nodata         &\nodata
  &  \nodata  &  \nodata  &   \nodata  &  \nodata \\
  L234E-N&16~48~05.6 &-10~54~29&$<$ 0.21&  \nodata &  \nodata        &
\nodata        &   \nodata        &   \nodata
  &  \nodata  &  \nodata  &   \nodata  &  \nodata \\
  L234E-S&16~48~08.7 &-10~57~25&0.62  &  \nodata  & 3.20 $\pm$ 0.01 &  2.89
$\pm$ 0.01 & 2.88 $\pm$ 0.01  & 3.08$\pm$0.01
  & 0.34 $\pm$ 0.02 & 0.353 $\pm$ 0.04 & -0.550 $\pm$ 0.06 &
-0.582$\pm$0.05\\
  L63&16~50~15.6 &-18~05~16&0.46    & 0.16 & 5.51 $\pm$ 0.01 &  5.46 $\pm$
0.01 & 5.48 $\pm$ 0.01  & 5.78$\pm$0.01
  & 0.24 $\pm$ 0.02 & -1.125 $\pm$ 0.11 & -1.350 $\pm$ 0.13 &
-1.267$\pm$0.12\\
  L462-1&18~07~31.3 &-04~43~29&0.26 &  \nodata   & 7.61 $\pm$ 0.02 &  7.72
$\pm$ 0.05 & 7.67 $\pm$ 0.05  & 7.89$\pm$0.01
  & 0.43 $\pm$ 0.04 & -0.660 $\pm$ 0.08 & -0.405 $\pm$ 0.13 & -0.505
$\pm$0.13\\
  L462-2\tablenotemark{a}&18~07~36.3 &-04~40~29&0.42  &  \nodata  & 7.89 $\pm$ 0.02 &  7.85
$\pm$ 0.02 & 8.07 $\pm$ 0.01  & 7.90$\pm$0.01
  & 0.35 $\pm$ 0.02 &-0.029 $\pm$ 0.06 &  -0.154 $\pm$ 0.07 &
0.471$\pm$0.05\\
  L429-1&18~17~05.7 &-08~13~30&0.39  &  \nodata  & 6.52 $\pm$ 0.01 &  6.45
$\pm$ 0.01 & 7.15 $\pm$ 0.01  & 6.85$\pm$0.01
  & 0.31 $\pm$ 0.03 & -1.074 $\pm$ 0.12 & -1.294 $\pm$ 0.13 &
  0.974$\pm$0.11\\
  L490&18~15~20.8 &-03~46~09&0.37    &  \nodata&  \nodata      &   \nodata
      & \nodata          &   \nodata
  & \nodata &  \nodata  &  \nodata &  \nodata\\
  L492&18~15~46.1 &-03~46~13&0.69    & 0.19 & 7.62 $\pm$ 0.01 &  7.60 $\pm$
0.01 & 7.61 $\pm$ 0.01  & 7.72$\pm$0.01
  & 0.52 $\pm$ 0.02 & -0.194 $\pm$ 0.03 & -0.240 $\pm$ 0.03 &
-0.208$\pm$0.03\\
  CB130-1&18~16~16.1 &-02~33~00&0.18  &  \nodata  & \nodata    &   \nodata
      & \nodata          &   \nodata
  &  \nodata  &  \nodata  &   \nodata  &  \nodata \\
  L328&18~17~09.0 &-18~02~01&$<$ 0.43&  \nodata&  \nodata        &   \nodata
        & \nodata          &   \nodata
  &  \nodata  &  \nodata  &   \nodata  &  \nodata\\
  L429-C&18~17~05.8 &-08~13~03&$<$ 0.58 &  \nodata&  \nodata        &
\nodata        &  \nodata         &   \nodata
  &  \nodata  &  \nodata  &   \nodata  &  \nodata \\
  L648-1&18~39~28.8 & 12~41~24&0.36 &  \nodata   & 4.00 $\pm$ 0.02 &  3.89
$\pm$ 0.04 & 3.69 $\pm$ 0.02  & 4.07$\pm$0.02
  & 0.29 $\pm$ 0.03 &  -0.269 $\pm$ 0.10 & -0.624 $\pm$ 0.18 & -1.303
$\pm$0.16\\
  L655&19~20~11.8 & 08~29~12&$<$ 0.27&  \nodata&  \nodata        &   \nodata
        &  \nodata         &   \nodata
  &  \nodata  &  \nodata  &   \nodata  &  \nodata \\
  L673-7&19~21~35.6 & 11~21~14&0.63  &  \nodata  & 6.93 $\pm$ 0.01 &  6.87
$\pm$ 0.01 & 6.84 $\pm$ 0.01  & 7.12$\pm$0.01
  & 0.54 $\pm$ 0.02 & -0.346 $\pm$ 0.03 & -0.467 $\pm$ 0.03 &
-0.520$\pm$0.03\\
  L694-2&19~41~04.6 & 10~57~02&0.17  &  $<$ 0.14& 9.40 $\pm$ 0.01 &  9.37
$\pm$ 0.01 & 9.33 $\pm$ 0.01  & 9.57$\pm$0.00
  & 0.27 $\pm$ 0.00 & -0.619 $\pm$ 0.03 & -0.756 $\pm$ 0.04 &
-0.874$\pm$0.04\\
  L1041-2&20~37~17.8 & 57~49~21&0.29 &  \nodata   &-1.97 $\pm$ 0.02 & -2.14
$\pm$ 0.01 &-2.00 $\pm$ 0.01  &-1.77$\pm$0.03
  & 0.39 $\pm$ 0.05 & -0.518 $\pm$ 0.11 & -0.941 $\pm$ 0.15 &
-0.600$\pm$0.11\\
  L922-1&20~38~29.8 & 44~32~00&0.28 &  \nodata   & 6.41 $\pm$ 0.01 &  6.43
$\pm$ 0.02 & 6.44 $\pm$ 0.02  & 6.61$\pm$0.01
  & 0.32 $\pm$ 0.03 & -0.637 $\pm$ 0.08 & -0.575 $\pm$ 0.10 &
-0.544$\pm$0.08\\
  L1049-1&20~41~04.9 & 57~27~31&0.30 &  \nodata   & 0.19 $\pm$ 0.03 &  0.23
$\pm$ 0.03 & 0.27 $\pm$ 0.03  & 0.13$\pm$0.02
  & 0.49 $\pm$ 0.03 & 0.124 $\pm$ 0.08 & 0.210 $\pm$ 0.07 & 0.284$\pm$0.07\\
  L1148&20~41~11.0 & 67~20~35&0.26  &  \nodata  & 2.37 $\pm$ 0.01 &  2.40
$\pm$ 0.02 & 2.37 $\pm$ 0.01  & 2.60$\pm$0.01
  & 0.29 $\pm$ 0.03 & -0.724 $\pm$ 0.09 & -0.728 $\pm$ 0.10 &
-0.741$\pm$0.11\\
  L1049-2&20~41~30.7 & 57~33~44&0.34 &  \nodata   &-0.15 $\pm$ 0.03 & -0.16
$\pm$ 0.03 &-0.17 $\pm$ 0.02  & 0.10$\pm$0.02
  & 0.41 $\pm$ 0.05 & -0.617 $\pm$ 0.11 &  -0.622 $\pm$ 0.11 &
-0.656$\pm$0.11\\
  L1155C2&20~42~58.9 & 67~48~18&0.20 &  \nodata   & 1.53 $\pm$ 0.02 &  1.49
$\pm$ 0.01 & 1.49 $\pm$ 0.01  & 2.68$\pm$0.01
  & 0.30 $\pm$ 0.03 & 0.527 $\pm$ 0.08 &  0.383 $\pm$ 0.06 &
0.400$\pm$0.06\\
  L1155C1&20~43~30.1 & 67~52~42&0.51 &  \nodata   & 2.52 $\pm$ 0.01 &  2.46
$\pm$ 0.02 & 2.49 $\pm$ 0.01  & 2.68$\pm$0.01
  & 0.44 $\pm$ 0.33 & -0.366 $\pm$ 0.28 & -0.509 $\pm$ 0.38 &
-0.432$\pm$0.33\\
  L981-1&21~00~13.2 & 50~20~50&$<$ 0.13  &  \nodata&  \nodata        &
\nodata        & \nodata          &   \nodata
  &  \nodata  &  \nodata  &   \nodata  &  \nodata \\
  L1063&21~08~10.3 & 56~18~20&$<$ 0.30 &  \nodata&  \nodata        &
\nodata        &   \nodata        &   \nodata
  &  \nodata  &  \nodata  &   \nodata  &  \nodata \\
  L944-2&21~17~46.9 & 43~18~20&0.37   &  \nodata & \nodata  &  \nodata &
\nodata  & 5.35$\pm$0.04
  & 0.34 $\pm$ 0.05 & \nodata  & \nodata & \nodata \\
  L1251A-2&22~27~41.6 & 75~13~31&0.36 &  \nodata   &-4.13 $\pm$ 0.03 & -4.31
$\pm$ 0.02 &-4.32 $\pm$ 0.01  &-3.97$\pm$0.02
  & 0.51 $\pm$ 0.03 & -0.306 $\pm$ 0.07 & -0.663 $\pm$ 0.07 &
-0.682$\pm$0.06\\
  L1197&22~37~02.4 & 58~57~21&0.26    & 0.06 &-3.34 $\pm$ 0.01 & -3.35 $\pm$
0.01 &-3.30 $\pm$ 0.01  &-3.16$\pm$0.01
  & 0.28 $\pm$ 0.03 & -0.629 $\pm$ 0.08 & -0.686 $\pm$ 0.09 &
-0.489$\pm$0.08\\
  L1251&22~37~31.3 & 75~10~23&0.23    &  \nodata&-3.71 $\pm$ 0.05 & -3.79
$\pm$ 0.04 &-3.62 $\pm$ 0.05  &-3.55$\pm$0.02
  & 0.56 $\pm$ 0.04 & -0.291 $\pm$ 0.10 & -0.430 $\pm$ 0.09 &
-0.123$\pm$0.09\\
  L1234&23~17~57.6 & 62~26~39&$<$ 0.27  &  \nodata&  \nodata        &
\nodata        & \nodata          &   \nodata
  &  \nodata  &  \nodata  &   \nodata  &  \nodata \\
  CB246-2\tablenotemark{a}&23~56~49.2 & 58~34~29&0.42   &  \nodata &-1.00 $\pm$ 0.01 & -1.05
$\pm$ 0.03 &-1.03 $\pm$ 0.02  &-0.84$\pm$0.01
  & 0.29 $\pm$ 0.03 & -0.548 $\pm$ 0.07 & -0.734 $\pm$ 0.12 &
-0.638$\pm$0.10 \\
\enddata
\tablecomments{ The first column indicates the name of the
observed region, and the second and third columns list the
coordinates observed toward the sources which are from the LM99
catalog. The 4th and 5th columns give antenna temperature of HCN
$J = 1-0$ ( $F$=2$-$1 ) and $\rm H^{13}CN$ $J=1-0$ lines. The
intensity values for sources with no detection in HCN or $\rm
H^{13}CN$ are given by 3 $\sigma$. The 6th, 7th, and 8th columns
are the velocities of the peak intensities of the HCN hyperfines.
The velocities and FWHMs of N$_{2}$H$^{+}$ spectra in columns 9
and 10 were taken from LMT99 by slightly shifting them from the
original values using the new frequency of N$_{2}$H$^{+}$ $J=1-0$
 taken from LMT01. The normalized velocity differences, $\delta$
V$_{HCN}$, given in the last three columns are calculated by e.q.
(1). } 
\tablenotetext{a}{These sources have double peaks in HCN spectra 
whose intensity ratios are close to 1.0 within error. The HCN
peak velocities for such sources were obtained by the Gaussian fit
after masking a fainter peak between two components. The $\delta$V values
of these sources are not used in our statistical analysis of $\delta$V }

\end{deluxetable}
\end{landscape}